\documentclass[]{aa}

\usepackage{graphicx}

\usepackage{txfonts}

\usepackage[]{hyperref}
\usepackage{caption}
\usepackage{subcaption}
\usepackage{placeins} 
\usepackage{gensymb}

\usepackage[version=4]{mhchem} 
\usepackage{bm} 
\usepackage[dvipsnames]{xcolor}

\begin{document} 

    \title{The Circumbinary Disc of HD\,34700A}
    \subtitle{II. Analysis of a strong dust asymmetry}

   \author{D.~Fasano
           \inst{\ref{inst1},\ref{inst2}}
          \and
          M.~Benisty
          \inst{\ref{inst2}}
          \and
          J.~Stadler
        \inst{\ref{inst1},\ref{ESO}}
          \and                 
          F.~Zagaria
          \inst{\ref{inst2}}
          \and 
          A.~Ziampras
          \inst{\ref{inst2},\ref{LMU}}
          \and 
          A.~Winter
          \inst{\ref{inst3}}
          \and 
          J.~Bae 
          \inst{\ref{inst4}}
          \and 
          S.~Facchini 
          \inst{\ref{inst5}}
          \and 
          N. T.~Kurtovic 
          \inst{\ref{inst2}}
          \and 
          E.~Ragusa 
          \inst{\ref{inst5}}
          \and 
          R.~Teague 
          \inst{\ref{inst6}}
          }

   \institute{Université Côte d'Azur, Observatoire de la Côte d'Azur, CNRS, Laboratoire Lagrange\\
              \email{daniele.fasano@oca.eu}\label{inst1}
        \and
            Max-Planck-Institut für Astronomie, Königstuhl 17, 69117 Heidelberg, Germany\label{inst2}
        \and
        European Southern Observatory, Karl-Schwarzschild-Str. 2, 85748 Garching bei München, Germany\label{ESO}
        \and
        Astronomy Unit, School of Physics and Astronomy, Queen Mary University of London, London E1 4NS, UK\label{inst3}
        \and
        Department of Astronomy, University of Florida, Gainesville, FL 32611, USA\label{inst4}
        \and
        Dipartimento di Fisica, Università degli Studi di Milano, Via Celoria 16, Milano, Italy\label{inst5}
         \and
         Department of Earth, Atmospheric, and Planetary Sciences, Massachusetts Institute of Technology, Cambridge, MA 02139, USA\label{inst6}
         \and
         Ludwig-Maximilians-Universit{\"a}t M{\"u}nchen, Universit{\"a}ts-Sternwarte, Scheinerstr.~1, 81679 M{\"u}nchen, Germany\label{LMU}
            }

   \date{Received November 21, 2025; accepted March 25, 2026}

  \abstract{ALMA observations have shown that substructures are ubiquitous in protoplanetary discs. A sub-group, the transition discs, shows large cavities and rings in dust continuum. Among these, some present very high contrast asymmetries possibly due to the presence of vortices. HD\,34700A is a binary system featuring a cavity, a ring, and multiple spiral arms detected in scattered light, a prominent crescent in the ALMA continuum and a complex gas morphology possibly connected with ongoing infall.}
    {We present new ALMA band 6 ($1.3~\rm mm$) continuum images of the circumbinary disc around HD\,34700A and compare them with two other systems showcasing high ($\gtrsim30$, measured as the peak-to-azimuthal-average ratio) contrast continuum asymmetries, IRS\,48 and HD\,142527. We aim to characterise the crescent morphology, discuss their possible origin, and, in the case of the vortex scenario, assess the efficiency of dust trapping in these systems.}
    {We perform visibility modelling of the new high resolution ($0\farcs11\times0\farcs09$) ALMA band 6 continuum data of HD\,34700A, together with improved visibility modelling of the other two targets.}
 {We detect a $0\farcs46$ ($161~\rm au$) large cavity and resolve a ring with an asymmetric crescent and an extended tail at $0\farcs53$ ($186~\rm au$) with peak intensity of $1.9~\rm mJy beam^{-1}$, corresponding to the second highest contrast ($\sim$62) ever detected with ALMA in a protoplanetary disc. We also detect unresolved emission inside the cavity, that we attribute to an  inner disc. Our visibility model is in remarkable agreement with the HD\,34700A data, featuring only localised residuals in the region of the disc corresponding to the tail of the asymmetry. For HD\,142527, we obtain a very good overall agreement with the data, recovering both the double peaked asymmetric ring and the inner disc emission. In the case of IRS\,48 we recover the general morphology of the asymmetry, but we cannot reproduce the fainter ring. We then run a hydrodynamic model of a vortex with different dust fluids, reproducing the general morphology observed in the HD\,34700A and IRS\,48 systems, with the emission around the vortex showing a mild asymmetry between the leading and trailing sides.}
    {With a combination of visibility, dust evolution and hydrodynamical models, we have constrained the morphology of the dust continuum emission of HD\,34700A for the first time, and improved existing models for IRS\,48 and HD\,142527. The high azimuthal contrast of the asymmetries rules out the orbit clustering of eccentric cavities scenario, while the dust evolution models we consider suggest that the vortex scenario is a plausible option.}

   \titlerunning{Analysis of a strong dust asymmetry in the circumbinary disc of HD\,34700A}
   \authorrunning{Fasano et al.}
   \maketitle

\section{Introduction}
\label{sec:Introduction}

High spatial resolution observations from the Atacama Large Millimeter/submillimeter Array (ALMA) have shown a variety of substructures in the continuum emission of protoplanetary discs, such as rings, gaps and cavities \citep{Andrews_2018, LongF_2018, Andrews_2020, Bae_2023}. Systems featuring large cavities in the dust emission are typically labelled as transition discs \citep{Strom1989,Espaillat2014}. Kinematic analysis of molecular line emission shows the presence of cavities also in the gas in many transition discs \citep{vdM2016b, Boehler2017,Dong2017}, as well as a more complex morphology showing spiral arms and localised deviations from Keplerian rotation \citep{Wolfer_2023, Izquierdo_2025}.

A systematic study of asymmetric transition discs was conducted by \citet{vdM2021}. More recently, similar continuum asymmetries have been detected also in debris discs, suggesting that they might be common also in the later stages of protoplanetary discs \citep{Marino_2026, Lovell_2026}. These systems are of particular interest as the asymmetric emission is thought to be caused by physical processes that promote azimuthal trapping of dust grains \citep{Birnstiel2013,Pinilla_2012}, and could potentially provide favourable conditions for the formation of planetesimals \citep{Barge_Sommeria_1995, Surville_2016, Surville_Mayer_2019}. Azimuthal dust trapping is primarily caused by the presence of a pressure maximum, with two possible mechanisms at its origin: anti-cyclonic vortices triggered by the Rossby Wave Instability (RWI), \citep{Lovelace1999, Li2000, Lyra2013, Baruteau2016} or gas horseshoes induced by eccentric cavities due to binary companions \citep{Shi2012,Ragusa2017,Miranda_2017}. The Rossby Wave Instability requires a narrow dip in the vortensity radial profile, which can occur in the presence of strong density \citep{deValBorro2007} or viscosity gradients \citep{Regaly2012}, typically at the edge of gas rings and cavities, and can be triggered by infalling material on the disc \citep{Bae2015}. It also requires low viscosity \citep[$\alpha\lesssim10^{-3}$][]{Shakura_Sunyaev_1973,Zhu2014} to produce long lived vortices. On the other hand gas horseshoes only need a massive binary companion (with a mass ratio of $q>0.05$) and are not limited by the disc viscosity \citep{Ragusa2017,Ragusa2020}. Both mechanisms produce a gas azimuthal overdensity of a factor $\lesssim2-4$, orbiting at Keplerian velocity, and trap millimetre dust grains both radially and azimuthally \citep{Birnstiel2013}, resulting in an azimuthal dust contrast, defined as the ratio between the peak intensity and the azimuthal average of the ring emission\footnote{For very high contrast discs, where the emission of the ring is not detected across all azimuth, a different approach is typically used, evaluating the contrast as the ratio between the peak intensity and either the emission at the same radius shifted by $180\degree$, or the $3\sigma$ upper limit.}, of up to a few hundreds. 

Another possible origin for continuum asymmetries is the presence of an eccentricity gradient in the disc \citep{Ataiee2013}, where the eccentric orbits of dust and gas particles cluster along the gradient direction, which is typically the apocenter for eccentric cavities in transition discs \citep{Ragusa_2024}. In this scenario the overdensity remains fixed at the apocenter of the orbit, hence it does not rotate with Keplerian motion around the central star. As the overdensity is a consequence of the orbits clustering and eccentricity gradient of the disc, it also does not create azimuthal pressure variations and does not trap dust, achieving only a moderate azimuthal contrast of $\lesssim5$ \citep{Ataiee2013}.

Differentiating between these scenarios as the cause of the observed asymmetries has been a challenging endeavour. The disc turbulence, which governs if vortices can be excited, is poorly constrained \citep{Flaherty2020, Rosotti_2023}. As a result, observations achieving an angular resolution higher than $0\farcs1$ and a spectral resolution of $\sim100\rm m s^{-1}$ \citep[e.g.][]{Robert_2020, Huang_2018} are needed to detect the characteristic gas signature associated with vortices and disentangle the origin of the dust asymmetries, although beam smearing effects can produce similar patterns \citep{Boehler2021}. \citet{vdM2021} studied a sample of asymmetric transition discs, but could not rule out either process. Recently, \citet{Wolfer2025} constructed an analytical vortex model and performed hydrodynamic simulations to study four discs with azimuthal dust asymmetries from the exoALMA sample \citep{Teague_2025}, finding no unambiguous link between the kinematical patterns of the data and the continuum crescents, likely due to the complex underlying kinematical field.

In this paper, we present recent ALMA observations of the system HD 34700 AaAb (hereafter HD\,34700\,A) that  resolved a prominent dust asymmetry co-located with a tentative kinematic vortex signature \citep{Stadler_2026}. We focus on the continuum data of HD\,34700\,A and perform visibility modelling of the new ALMA Band 6 observations, comparing it with the transition discs with the most prominent asymmetries (with a contrast $\gtrsim30$) from the \citet{vdM2021} sample, namely Oph-IRS 48 (IRS\,48 hereafter) and HD\,142527. This paper is organised with the following structure: in Sec.~\ref{sec:Observations} we present the observations used in our analysis, while in Sec.~\ref{sec:Methods} we introduce the methods adopted to model the data. In Sec.~\ref{sec:Results} we present the results of our visibility modelling and discuss them in Sec.~\ref{sec:Discussion}; we summarise our findings in Sec.~\ref{sec:Conclusions}.

\section{Observations}
\label{sec:Observations} 

\subsection{Sources}
\label{sec:Sources}

\paragraph{\textit{HD\,34700\,A:}} HD\,34700\,A is a near-equal-mass ($M_{\rm bin}=4~M_\odot$) close ($a_{\rm bin}=0.69~\rm au$) Herbig Ae spectroscopic binary, belonging to the quadruple system HD 34700 \citep[$d\approx351~\rm pc$][]{Gaia2023, Sterzik2005}. The two young ($\sim5~\rm Myr$) stars have stellar radii of $R_1=3.46~R_\odot$ and $R_2=3.40~R_\odot$ \citep{Monnier2019}, and effective temperatures of $T_{\rm eff,1} = 5900~\rm K$ and $T_{\rm eff,1} = 5800~\rm K$ \citep{Torres_2004}. \citet{Monnier2019} resolved the circumbinary disc around HD\,34700\,A for the first time using Gemini Planet Imager (GPI) band K and H polarized scattered light images, unveiling an elliptical ring at $175~\rm au$ with an inclination $i=42\degree$ and a Position Angle (PA) of $42\degree$, a dust depleted cavity, multiple spiral arms and a discontinuity towards the north side of the disc, later confirmed by \citet{Uyama2020} with Subaru/SCExAO and \citet{Columba2024} using SPHERE/IRDIS (H and K bands) and SPHERE/IFS (YJH bands). Additionally, \citet{Columba2024} detected a symmetric ring with inclination $i=39.7\pm2\degree$ and $\rm PA=87.1\pm1.2\degree$ in SPHERE/ZIMPOL $\rm H\alpha$ scatter light observations, inside the infrared (IR) ring, at $65-120~\rm au$. \citet{Benac2020} imaged the $1.13~\rm mm$ continuum, $^{12}$CO and $^{13}$CO $\rm J=2-1$ emission with a resolution of $0\farcs5$ using the Submillimeter Array (SMA), resolving a prominent dust asymmetry at $155^{+11}_{-7}~\rm au$ with radial width of $72^{+14}_{-15}~\rm au$ and azimuthal width $64^{+8}_{-7}$~deg, and a circumbinary disc rotating counter-clockwise. 

Recent ALMA band 6 observations ($\lambda=1.3~\rm mm$, $0\farcs11$ resolution), \citep{Stadler_2026} resolved for the first time the ring co-orbital with the previously detected asymmetry, a dust depleted cavity and emission from the inner region in the continuum. We describe the dust morphology in detail in Sec.~\ref{sec:Results}. \citet{Stadler_2026} also performed kinematic analysis of the $^{12}$CO, $^{13}$CO and C$^{18}$O $\rm J=2-1$ molecular lines, tracing a $0\farcs20$ ($\approx65~\rm au$) inner cavity in the peak intensity maps of all the isotopologues, strong deviations from Keplerian rotation in the $^{12}$CO emission in the shape of spiral spurs following the IR spiral ring, and less pronounced residuals in the $^{13}$CO and C$^{18}$O velocity maps resembling the kinematic signature of a vortex \citep{Wolfer2025}. 

\paragraph{\textit{IRS\,48:}}
IRS\,48 is a $\sim2.5~M_\odot$, $\sim4~\rm Myr$ old Herbig A0 star \citep[$T_{\rm eff}\approx9000~\rm K$;][]{Brown2012, Schworer_2017, Francis2020}, at a distance of $d\approx135~\rm pc$ \citep{Gaia2023}. ALMA band 7 observations of $^{12}$CO and $^{13}$CO have shown a gas disc with inclination $i=50\degree$ and $\rm PA=100\degree$ and a $20-30~\rm au$ gas cavity \citep{Bruderer2014,vdM2016}. The continuum emission features an eccentric ring ($e=0.27$) with a semi-major axis of $78~\rm au$ and the most prominent asymmetry observed at sub-mm wavelengths (with a contrast of $\gtrsim270$) \citep{Yang2023}, interpreted as a dust trap \citep{vdM2013, vdm2015}. The asymmetry shows a tail spanning $\sim180\degree$ from the peak emission to the opposite side of the ring, that trails behind the overdensity, as the disc rotates counter-clockwise \citep{Bruderer2014}. Using polarimetric ALMA data, \citet{Yang2024} attributed the asymmetric overdensity to a vortex with moderate dust settling and an effective turbulence parameter $\alpha=10^{-4}-5\times10^{-3}$.

\paragraph{\textit{HD\,142527:}}
HD\,142527 is a binary system comprising a $\sim1.69~M_\odot$ F6 star, with an age of $\sim8~\rm Myr$, stellar radius of $R\approx2.2~\rm au$ and temperature of $\sim6500~\rm K$ \citep{Fairlamb2015,Francis2020}, and a $\sim0.1~M_\odot$ M dwarf companion \citep[$T_{\rm eff}=2600-2800~\rm K$, $R\approx0.9~R_\odot$;][]{Lacour_2016, Claudi2019} on an eccentric ($e\gtrsim0.2$) and non coplanar orbit, at separations of $\sim44-90~\rm mas$ \citep[$\sim15~\rm au$;][]{Biller2012, Nowak2024} and a distance of $d\approx157~\rm pc$ \citep{Gaia2023}. The binary orbit is too compact to explain the large cavity, suggesting the presence of a planet shaping the disc substructures \citep{Nowak2024}. Recent MATISSE observations have shown that the circumstellar disc around HD\,142527\,A features a complex, asymmetric geometry, possibly associated with the interactions with the companion star \citep{Scheuck_2026}. The stars are surrounded by a circumbinary disc rotating clockwise with inclination $i=27\degree$ \citep{Fukagawa2013, vdM2021} and a position angle $\rm PA=160\degree$. A double peaked asymmetry at a radial location of $\sim166~\rm au$ has been observed in the ALMA band 7 continuum, with a contrast of $\gtrsim50$ \citep{Casassus2013, Boehler2017}. The $^{12}$CO and C$^{18}$O ALMA line emission have shown a gas cavity of $50~\rm au$, wherein a misaligned inner disc \citep[$i=23\degree$, $\rm PA=14\degree$][]{Marino2015,Bohn2022} casts shadows on the outer disc \citep{Boehler2017}. Performing kinematic analysis of these molecules, \citet{Boehler2021} found deviations of $350~\rm m s^{-1}$ from the Keplerian rotation of the disc, co-located with the dust crescents, and suggesting that this could be caused by a vortex with an aspect ratio of 5, centred at the secondary peak of the dust continuum. Additionally, \citet{Temmink2023} analysed multiple molecules in this system, observing an asymmetric molecular emission in HCN and CS, peaking at the opposite side of the disc with respect to the continuum crescent. They infer that this asymmetry is likely due to continuum over-subtraction for the main CO isotopologues, and propose that it can be the result of shadowing from the misaligned inner disc for the CS and HCN molecules.

\subsection{Data}
\label{sec:Data}

In this work, we consider ALMA band 6 continuum data of the system HD\,34700\,A (2022.1.00760.S; PI: J. Stadler), ALMA band 7 ($0.87~\rm mm$) continuum observations of the disc IRS\,48 (2019.1.01059.S; PI: H. Yang) and ALMA band 7 continuum data of the system HD\,142527 (2013.1.00305.S; PI: S. Casassus). 

The HD 34700 data consists of two short baseline (SB) and 11 long baseline (LB) execution blocks (EBs). The SB data were taken on October 9, 2022, in configuration C-3 (maximum baseline $L_{\rm max}=500~\rm m$) and with a mean Precipitable Water Vapour (PWV) of $0.3~\rm mm$. The LB data were taken in configuration C-6 ($L_{\rm max}=2.5~\rm km$) over a period of three weeks, on May 13, 19, 21, 24, 28, 30, and June 2, 2023, using a number of antennas between 42 and 44 and with a mean PWV in the range $0.3-1.4~\rm mm$. The spectral setup was chosen to study the kinematics of the system, thus only two spectral windows (SPWs) are dedicated to the continuum emission centred at $218.0$ and $232.6~\rm GHz$ with the maximum bandwidth of $1.9~\rm GHz$. 
In this paper, we focus on visibility modelling the continuum emission of HD\,34700\,A. However, the HD~\,34700 ALMA data set incorporates emission not only from the circumbinary disc A, but also from the low-mass companion B in the same field-of-view \citep{Stadler_2026}. We first start by removing the contribution from the secondary from the visibility table. To do so, we create a CLEAN model of HD\,34700\,B CLEANing down to $5\mu\rm{Jy/beam}\,(\approx 1\sigma)$ within an elliptic CLEAN mask of $R=0\farcs625$ ($i=41.5$) centred around B. We then remove the \texttt{tclean} model visibility from the data set using the \texttt{uvsub()} task of CASA \citep{CASA_2022} to obtain the final measurement set used in this work. The phase centre of the dataset is $(05^h19^m41^s.4115, 05\degree38'42\farcs7931)$. 

The IRS\,48 observations consists of six EBs in two different array configurations, C43-6 and C43-7, and were conducted on 2021 June 7, June 14, and July 19 with a mean PWV in the range $0.6-0.9~\rm mm$. As the observations were aimed at measuring the dust polarization, all the SPWs were dedicated to the continuum emission, centred at $336.5$, $338.4$, $348.5$, and $350.5~\rm GHz$, with a nominal bandwidth of $2.0~\rm GHz$. The reported phase centre is $(16^h27^m37^s.1700, -24\degree30'35\farcs6710)$.

The HD\,142527 data consists of two EBs in two different array configurations, C43-5 and C35-5. The observations were conducted on August 11 and 30, 2015, with a PWV of $1.6~\rm mm$ and only one spw dedicated to the continuum emission, centred at $277~\rm GHz$ with a nominal bandwidth of $2.0~\rm GHz$. The reported phase centre is $(15^h56^m41^s.8741, -42\degree19'23\farcs6564)$.

The self-calibration strategy for the HD\,34700\,A, IRS\,48 and HD\,142527 datasets are described in \citet{Stadler_2026, Yang2023,Temmink2023}.

\subsection{Imaging}
\label{sec:Imaging}

We image the continuum data using the \texttt{tclean()} task of the CASA software \citep{CASA_2022}, version 6.6.5-31. We define elliptic masks with semi-major axis of $1\farcs2$, $1\farcs0$, $2\farcs2$ and inclination of $35\degree$, $51\degree$, $29\degree$ for HD\,34700\,A, IRS\,48, and HD\,142527, respectively, and CLEAN the images down to $3\sigma$ using a robust parameter of $0.0$, chosen to provide the best balance between resolution and sensitivity. We produce a $2000\times2000$ pixels image with pixel size of $0\farcs015$ for HD\,34700\,A, a $1024\times1024$ pixels image with pixel size of $0\farcs009$ for IRS\,48, and a $1024\times1024$ pixels image with pixel size of $0\farcs025$ for HD\,142527. We use the \texttt{gofish} package \citep{GoFish} to compute the Root Mean Square (RMS) noise in annulus between $2\farcs5-4\arcsec$ for HD\,34700\,A and IRS\,48, and between $3\arcsec-4\arcsec$ for HD\,142527. The final images are shown in Figure~\ref{fig:Continuum gallery}, while their properties are summarised in Table~\ref{table:Disc continuum properties}.

    \begin{table*}
    \caption{Disc continuum properties and geometrical parameters.}            
    \label{table:Disc continuum properties}     
    \centering        
    \footnotesize
    \renewcommand{\arraystretch}{1.5}
    \begin{tabular}{c c c c c c c c c c}        
    \hline\hline                
    System & Robust, Beam & PA$_{\rm beam}$ & RMS noise & Peak Intensity & Contrast & $i$ & PA & $\Delta\rm RA$ & $\Delta\rm Dec$\\
     & [mas$\times$mas] & [deg] & [$\mu$Jy beam$^{-1}$] & [mJy beam$^{-1}$] &  & [deg] & [deg] & [mas] & [mas]\\ 
    \hline                      
       HD\,34700\,A & 0.0, 114 $\times$ 94 & 87 & 7 & $1.90\pm0.19$ & 62 & $35.09_{-0.06}^{+0.12}$ & $92.9_{-0.3}^{+0.3}$ & $-48_{-1}^{+1}$ & $14_{-1}^{+1}$\\    
       IRS\,48 & 0.5, 108 $\times$ 72 & $-73$ & 16 & $18.0\pm1.8$ & 259 & $50.996_{-0.005}^{+0.003}$ & $100.12_{-0.04}^{+0.04}$ & $83.8_{-0.3}^{+0.3}$ & $107.8_{-0.2}^{+0.2}$\\
       HD\,142527 & 0.0, 299 $\times$ 203 & 79 & 252 & $85.8\pm8.6$ & 37 & $29.079_{-0.004}^{+0.008}$ & $182.96_{-0.03}^{+0.05}$ & $-38.1_{-0.2}^{+0.2}$ & $-154.6_{-0.2}^{+0.2}$\\
    \hline                                   
    \end{tabular}

    \end{table*}

\section{Methods}
\label{sec:Methods} 

We model the continuum emission of the new HD\,34700\,A data in order to constrain the morphological properties of the crescent and ring. \citet{vdM2021} already fitted the visibilities of the ALMA data for the systems IRS\,48 and HD\,142527, assuming a simple 2D arc to model the asymmetric rings. However, recent observations of IRS\,48 also resolve a ring, while a single arc could not fully describe the double peaked emission of HD\,142527, as shown by their intensity residuals of up to $40\%$. As a result, in this work we perform visibility fitting for the new data of HD\,34700\,A as well as the IRS\,48 and HD\,142527 data, using an improved model. 
 
 We use the code \texttt{galario} \citep{Tazzari_2018} to fit the visibility data and characterize the morphological structure of the continuum emission. We assume a 2D parametric intensity model and follow a Markov Chain Monte Carlo (MCMC) approach implemented with the package \texttt{emcee} \citep{Foreman-Mackey_2013}. With this procedure, we can constrain both the morphological parameters specific to the chosen model, as well as the geometrical parameters of the outer disc, namely the inclination $i$, the position angle PA and the offsets $(\Delta\rm RA, \Delta Dec)$ between the disc and the phase centres. 

We define three models for the brightness distribution:

\begin{align}\label{eq:Intensity HD34700}
I_{\rm HD~34700~A}(r,\theta) &= G(f_0, \sigma_{0}) + GR(f_1, r_1, \sigma_{r,1}) \nonumber \\
&\quad + GAa(f^{\rm a}_1, r^{\rm a}_1, \sigma^{\rm a}_{r,1}, \theta^{\rm a}_1, \sigma^{\rm a}_{\theta,\rm l1}, \sigma^{\rm a}_{\theta,\rm r1}),
\end{align}

\begin{align}\label{eq:Intensity IRS48}
I_{\rm IRS~48}(r,\theta) &= GR(f_1, r_1, \sigma_{r,1}) \nonumber \\
&\quad + GAa(f^{\rm a}_1, r^{\rm a}_1, \sigma^{\rm a}_{r,1}, \theta^{\rm a}_1, \sigma^{\rm a}_{\theta,\rm l1}, \sigma^{\rm a}_{\theta,\rm r1}),
\end{align}

\begin{align}\label{eq:Intensity HD142527}
I_{\rm HD~142527}(r,\theta) &= Go(f_0, x_0, y_0, \sigma_{0}) + GR(f_1, r_1, \sigma_{r,1}) \nonumber \\
&\quad + \sum_{i=1}^{2} GAa(f^{\rm a}_i, r^{\rm a}_i, \sigma^{\rm a}_{r,i}, \theta^{\rm a}_i, \sigma^{\rm a}_{\theta,\mathrm{l}i}, \sigma^{\rm a}_{\theta,\mathrm{r}i}),
\end{align}
where $G(f, \sigma)$ is a Gaussian aligned to the image centre with radial width $\sigma$ and peak flux $f$, $Go(f, x, y, \sigma)$ is an offset Gaussian centred in the $(x,y)$ coordinates, $GR(f, r, \sigma)$ is an axisymmetric Gaussian ring centred in $r$ with radial width $\sigma$ and peak flux $f$, while $GAa(f, r, \sigma_r, \theta, \sigma_{\theta,\rm l}, \sigma_{\theta,\rm r})$ is a Gaussian asymmetric arc centred in $(r,\theta)$ with radial width $\sigma_r$, azimuthal widths $\sigma_{\theta,\rm l}$ and $\sigma_{\theta,\rm r}$, and peak flux $f$. The models for HD\,34700\,A, IRS\,48 and HD\,142527 have a total of $15$, $13$ and $23$ parameters, respectively. We use the Gaussian to describe the emission originating inside the cavity, the Gaussian ring, and the Gaussian arcs to model the asymmetric ring, choosing one arc for the single peaked emission of HD\,34700\,A and IRS\,48, and two arcs for the double peaked morphology of HD\,142527. We use asymmetric arcs with different azimuthal widths on the left ($\sigma_{\theta,\rm l}$) and right ($\sigma_{\theta,\rm r}$) of the peak intensity to better reproduce the tail observed in the continuum images. We choose to fix the inner Gaussian to the image centre for the HD\,34700\,A model due to its proximity to the ring asymmetry, as the inner Gaussian would merge with the ring otherwise. We check that fixing the Gaussian in the centre does not affect the remaining parameters of the model, as they remain within $1\sigma$ error bars when the Gaussian is not fixed in the centre. In the case of IRS\,48, adding a Gaussian component to the model does not recover the emission from the inner disc. This might be caused by the extreme intensity contrast of the asymmetric emission. As a result, we choose not to use a Gaussian component in the model of IRS\,48. 

We perform a preliminary exploration of the parameter space with \texttt{emcee} using uniform priors, sampling $f_0$, $f_1$, $f_1^a$ and $f_2^a$ logarithmically and the remaining parameters linearly. As initial step we use 100 walkers over $\sim10^3$ steps. We then associate the best fit parameters as the median of the posterior distribution from this preliminary exploration, and set them as initial guesses for the fiducial \texttt{emcee} run. For each dataset we set up 120 walkers and $2\times10^4$ steps, and discard the first $25\%$ steps as burn-in. We report the geometrical parameters ($i$, PA, $\Delta\rm RA$ and $\Delta\rm Dec$) in Table~\ref{table:Disc continuum properties}, while the full set of best fit parameters is shown in Table~\ref{table:galario results}, with the adopted priors reported in Appendix~\ref{App: Visibility Modelling}. For the inclination and PA of HD\,34700\,A, we consider narrow priors around the values resulting from the $^{13}$CO \texttt{discminer} \citep{Izquierdo_2021} fit in \citet{Stadler_2026} to help achieve convergence, and be consistent with their analysis. We use the same procedure for IRS\,48, using the inclination and PA reported in \citet{Yang2023}. For HD\,142527, instead, we setup wider priors around the inclination and PA reported in \citet{vdM2021}. We then evaluate the best fit \texttt{galario} model sampling the visibility plane at the same $uv$ points of the observations, obtaining synthetic visibilities for the model, and subtract them from the observed visibilities in order to obtain the residuals. We then image the best fit \texttt{galario} model and the residuals using the same procedure explained in Sec.~\ref{sec:Imaging}.

\section{Results}
\label{sec:Results} 

\subsection{Continuum images}
\label{sec:Continuum images} 

In Figure~\ref{fig:Continuum gallery} we present the continuum image of the new ALMA band 6 data for HD\,34700\,A and compare it with the continuum images of IRS\,48 \citep{Yang2023} and HD\,142527 \citep{Temmink2023}. In Figure~\ref{fig:Azimuthal plots}, we show the de-projected continuum images in polar coordinates (left panel) and azimuthal profiles at $\{87, 112, 137, 162, 187\}~\rm au$, $\{65, 80, 95, 110, 125\}~\rm au$ and $\{137, 162, 187, 212, 237\}~\rm au$ for HD\,34700\,A, IRS\,48, HD\,142527, respectively. We perform the de-projection using the package \texttt{gofish}, assuming the geometrical parameters from Table~\ref{table:Disc continuum properties}, and then interpolate the resulting image on a $501\times2001$ ($R, \theta$) grid, with $R\in[0, 4]\arcsec$ and $\theta\in[-180,180]\degree$.

    \begin{figure*}[!h]
    \centering
    \includegraphics[width=0.9\textwidth]{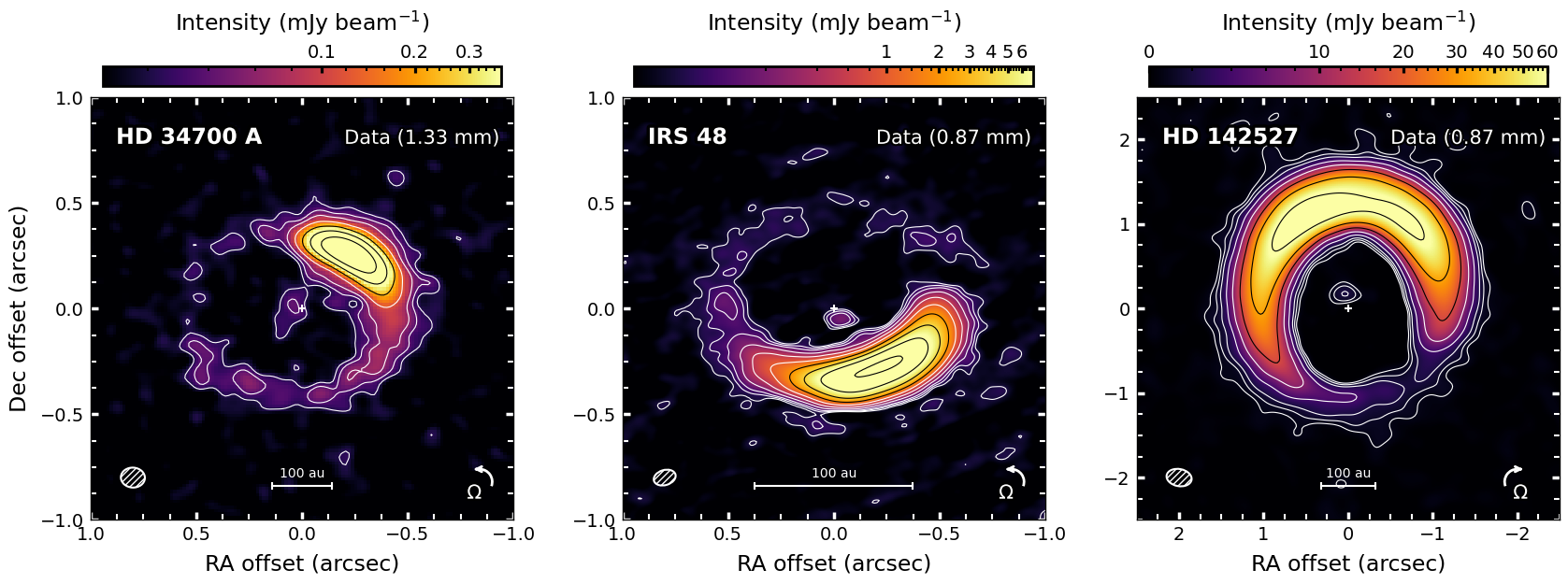}
     \label{fig:Continuum gallery}
     \caption{Continuum images of HD\,34700\,A (left), IRS\,48 (middle) and HD\,142527 (right). The white/black contours are taken at $3\sigma$, $5\sigma$ and $2^n\sigma$, with integer numbers $n\geq3$. The white plus sign marks the centre of the ring in the \texttt{galario} model, the ellipse in the bottom left corner represents the synthesised beam and the arrow in the bottom right corner shows the direction of the gas rotation. We apply an asinh stretch from $\{0.007, 0.007, 0.000\}~\rm mJy beam^{-1}$ to a factor $\{0.2, 0.4, 0.7\}$ of the peak intensity, using a stretch parameter of $\{0.1, 0.01, 0.1\}$ to the colour scale to visually enhance the fainter emission.}
     \label{fig:Continuum gallery}
    \end{figure*}

    \begin{figure*}[!h]
    \centering
    \includegraphics[width=0.9\textwidth]{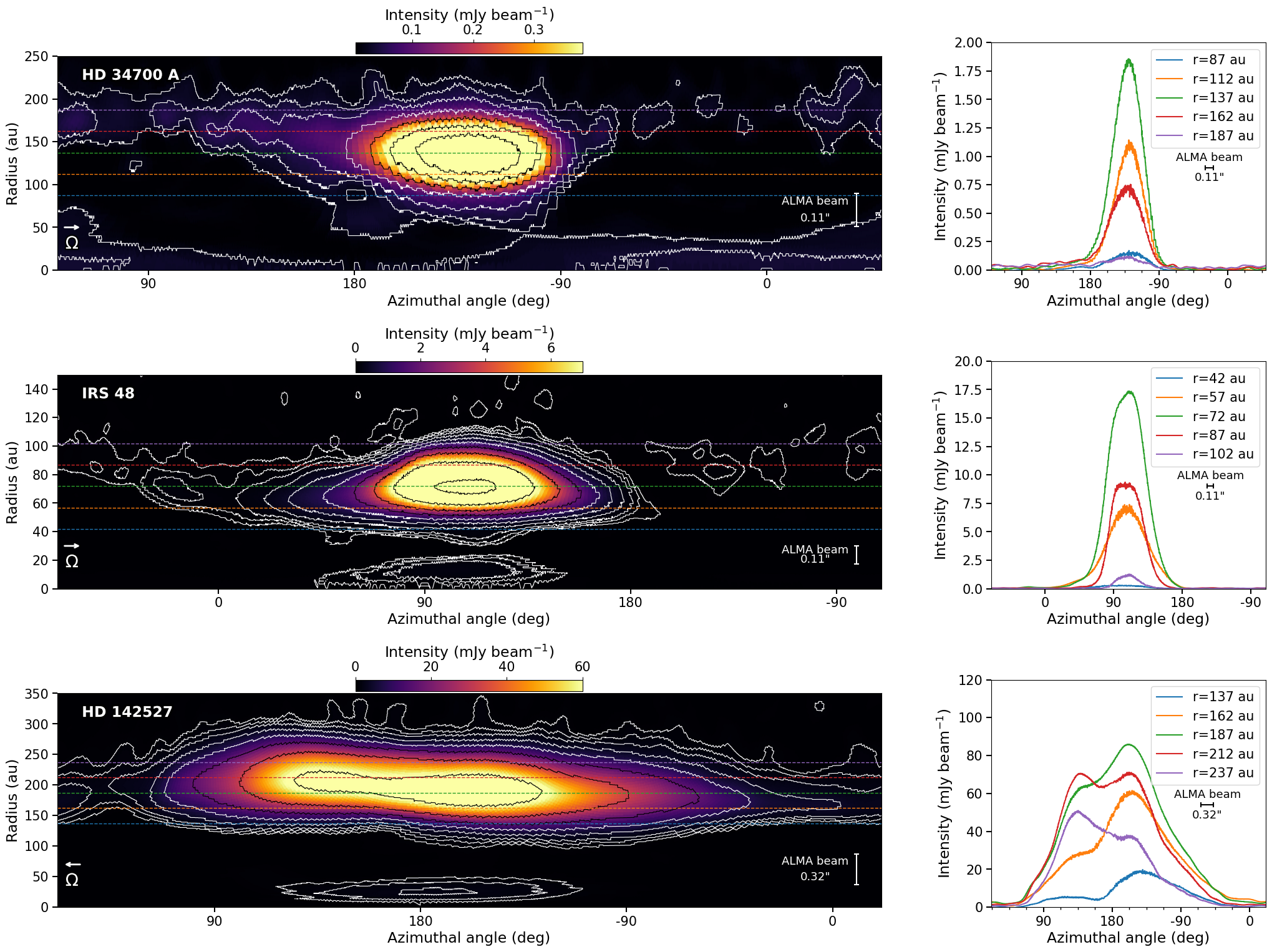}
     \label{fig:Azimuthal plots}
     \caption{Left column: de-projected continuum images of HD\,34700\,A (top), IRS\,48 (middle) and HD\,142527 (bottom) in polar coordinates. Contours are the same as in Figure~\ref{fig:Continuum gallery}. In the bottom left corner we show the rotation direction of the gas, while in the bottom right corner we show the beam size in the radial direction. Right column: Azimuthal profiles of the de-projected continuum images from the left column, taken at different radii around the peak location, corresponding to the dashed lines with the same colours in the left column. We use the \texttt{savgol\_filter} function of the \texttt{scipy} \citep{Virtanen_2020} module to smooth the oscillations introduced by the interpolation on the polar grid. We also show the ALMA beam size associated to the azimuthal profile evaluated at the peak radius. The plots have been shifted azimuthally so that the image peak is in the centre and the $-180\degree$ azimuth coincide with the $180\degree$ azimuth.}
     \label{fig:Azimuthal plots}
    \end{figure*}

All systems feature an asymmetric overdensity with a tail trailing with respect to the disc rotation, and emission originating inside the dust depleted cavity. In Figure~\ref{fig:Continuum gallery} we also show the centre of the ring in the \texttt{galario} model. We measure the peak intensity for all the images, estimating its uncertainty as the $10\%$ ALMA flux calibration error for band 6 and 7\footnote{see Sect. 10.2.6 in the \href{https:/almascience.nrao.edu/proposing/technical-handbook/}{ALMA Technical Handbook}}. We define the contrast as the ratio between the peak intensity and the azimuthal average of the ring emission\footnote{For HD\,142527 we compute the azimuthal average along the same radius of the peak intensity. For HD\,34700\,A and IRS\,48, instead, we perform the azimuthal average along the radius at $170~\rm au$ and $65~\rm au$, as the ring looks more eccentric.}, masking the asymmetry contribution when it becomes higher than $16\sigma$, and report their values in Table~\ref{table:Disc continuum properties}. This corresponds to a peak intensity of $271\sigma$, $1125\sigma$ and $340\sigma$ and contrasts of $62$, $259$ and $37$ for HD\,34700\,A, IRS\,48 and HD\,142527, respectively, making HD\,34700\,A the disc with the second most prominent contrast in continuum emission observed with ALMA to date. For reference, we also report a contrast of $45$, $365$ and $33$ if defined as the ratio between the peak intensity and the emission at the opposite side of the disc (i.e., same radius and PA shifted by $180\degree$). 

\subsection{\texttt{galario} model and residuals}
\label{sec:Galario model and residuals} 

   \begin{figure*}[!h]
    \centering
    \includegraphics[width=0.9\textwidth]{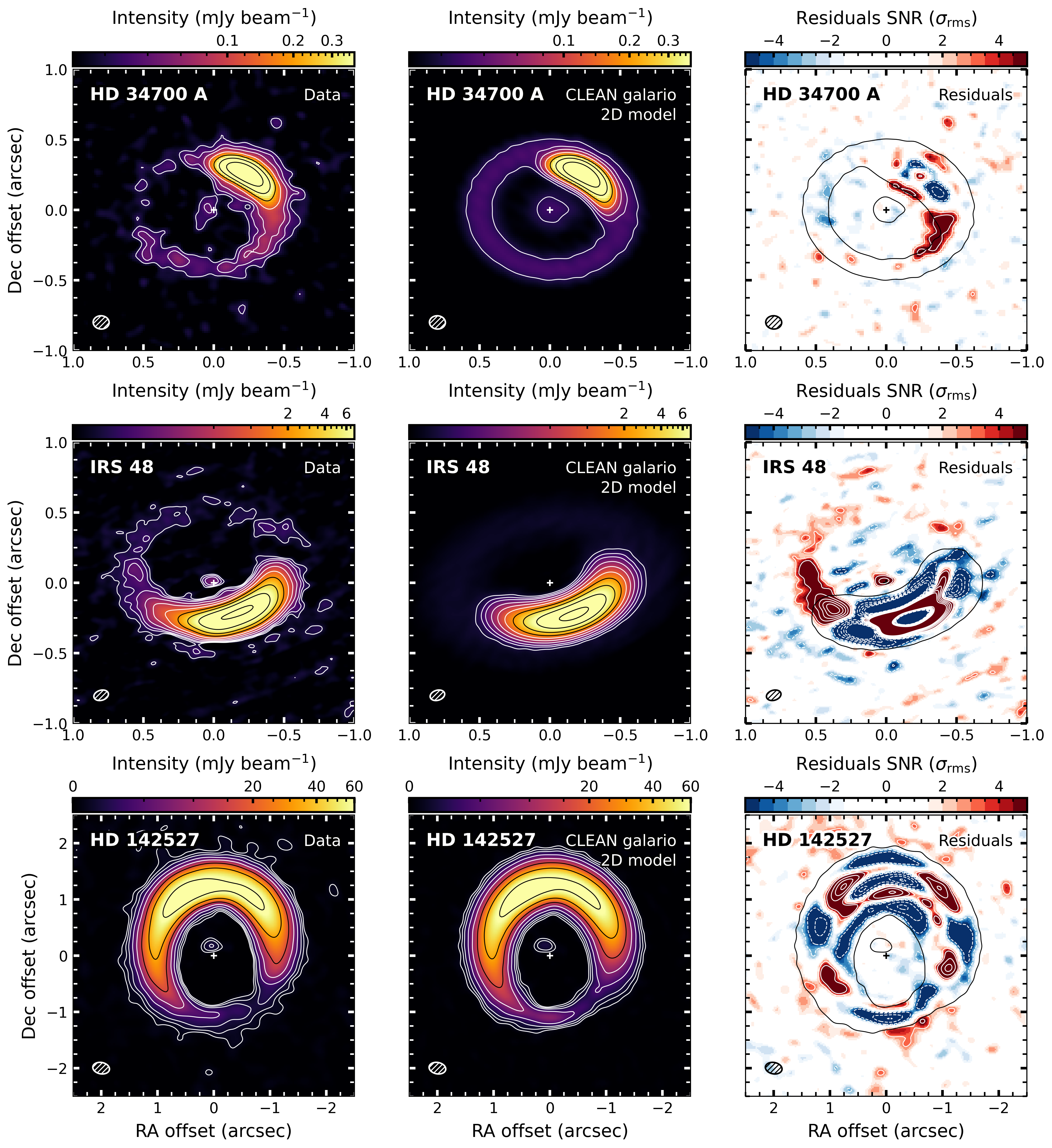}
     \label{fig:Summary plots}
     \caption{Top to bottom: Images of HD\,34700\,A; IRS\,48; HD\,142527. Left to right: Continuum images, same as in Figure~\ref{fig:Continuum gallery}; CLEANed \texttt{galario} model images; CLEANed residual images; In the left and middle columns, contours are the same as in Figure~\ref{fig:Continuum gallery}. In the right column, white solid (positive) and dashed (negative) contours start at $3\sigma$ and $5\sigma$, and increase by $5\sigma$, while black contours correspond to the $3\sigma$ emission from the CLEANed model.}
     \label{fig:Summary plots}
    \end{figure*}

In Figure~\ref{fig:Summary plots} we show the main results of the \texttt{galario} modelling of HD\,34700\,A (top panels), IRS\,48 (middle panels) and HD\,142527 (bottom panels). The \texttt{galario} model for HD\,34700\,A is in remarkable agreement with the data, with residuals of up to $7.4\sigma$ ($\sim2.7\%$ of the peak intensity) localised only in the region of the asymmetric tail. The final model consists of a central Gaussian with width of $\sim0\farcs13$ ($46~\rm au$), a ring centred in $\sim0\farcs53$ ($186~\rm au$) with radial width $\sim0\farcs08$ ($28~\rm au$) and an arc centred at $(\sim0\farcs39,328\degree)$ ($137~\rm au$) with radial width $\sim0\farcs03$ ($11~\rm au$) and an azimuthal Full Width Half Maximum (FWHM) of $43.2\degree$, which we computed as $\sqrt{2\ln{2}}(\sigma^a_{\theta,\rm l}+\sigma^a_{\theta,\rm r})$. Although we fixed the inner Gaussian to the disc centre, the model still reproduces the emission remarkably well, with residuals $<3\sigma$ in the cavity. The ring prescription we use is too simple to reproduce the faint azimuthal variations observed in the data, overestimating the intensity in the North-East side of the ring (with residuals $\lesssim3\sigma$) and underestimating the tail contribution, implying that the tail structure is more complex than a simple Gaussian azimuthal dependence. Taking into account the improved resolution and S/N, our results are consistent with the SMA data presented in \citet{Benac2020}, with their radial position, radial width and azimuthal FWHM of the asymmetry being within $2\sigma$ of our model fit and new data. 

For IRS\,48, the \texttt{galario} model reproduces the general morphology of the asymmetry, but fails to recover the ring emission. This is probably due to the extreme contrast of the crescent, which dominates the residuals, but can also be a consequence of the eccentric nature of the ring, which is not taken into account by our model. In the region of the asymmetry we measure residuals of up to $96\sigma$ and $-152\sigma$, which correspond to deviations of $8.6\%$ and $13.5\%$ of the peak emission, respectively. The final model consists of a ring centred in $\sim0\farcs72$ ($97~\rm au$) with radial width $\sim0\farcs17$ ($23~\rm au$), although too faint compared with the observations, and an arc centred at $(\sim0\farcs42,197\degree)$ ($56~\rm au$) with radial width $\sim0\farcs07$ ($9~\rm au$) and an FWHM of $79.0\degree$. These values are consistent with the single arc model from \citet{vdM2021}.

In the case of HD\,142527, the \texttt{galario} model reproduces the general morphology observed in the data at a very good qualitative level, showing both the disc and the double peaked emission in the ring. However, strong residuals of up to $20.1\sigma$ are present all across the ring, although they represent deviations of only $6.2\%$ when compared to the peak intensity, an improvement of more than a factor 6 compared to the previous model from \citet{vdM2021}. The final model features a Gaussian with an offset of ($94, -159$) mas and a width of $\sim0\farcs03$, a very thin ring centred at $\sim1\farcs1$ with width $0\farcs003$ and two asymmetric arcs centred at $(\sim1\farcs2,298\degree)$ and $(\sim1\farcs3,227\degree)$, with radial width of $\sim0\farcs14$ and $\sim0\farcs18$ and an azimuthal FWHM of $82\degree$ and $107\degree$, in contrast with the single arc centred at $240\degree$ with an FWHM of $155\degree$ modelled by \citet{vdM2021}. Additionally, we find an inclination and PA that differ by $2\degree$ and $23\degree$ from those reported in \citet{vdM2021}. 

\section{Discussion}
\label{sec:Discussion}

\subsection{Dust trapping}
\label{sec:Dust trapping}

In \citet{vdM2021}, the authors analysed a sample of transition discs showing a variety of crescents, aiming to model them as anticyclonic vortices. They constructed emission azimuthal profiles based on the analytical model\footnote{\url{https://github.com/birnstiel/azimuthal_profile}} of \citet{Birnstiel2013}, which computes analytically the equilibrium between azimuthal drift and mixing of dust particles, and comparing their FWHM with that extracted from the data azimuthal profiles, as a function of the observed Stokes number defined as:

\begin{equation}
    \mathrm{St}_{\rm obs} = \frac{\lambda_{\rm obs}\rho_{\rm s}}{4\Sigma_{\rm gas}(R_{\rm dust})},
\end{equation}
with $\lambda_{\rm obs}$ the observed wavelength, $\rho_{\rm s}$ the intrinsic dust density and $\Sigma_{\rm g}(R_{\rm dust})$ the gas density at the ring location $(R_{\rm dust})$. Then, they model the gas surface density distribution as

\begin{equation}\label{eq:azimuthal surf dens}
    \Sigma(R, y) = \Sigma_{\rm g}(R)\frac{1 + (A_{\rm gas}-1)\exp\left(-\frac{y^2}{2R^2\sigma_y^2}\right)}{1+(A_{\rm gas}-1)\frac{\sigma_y}{2\sqrt{2\pi}}\mathrm{Erf}\left(\frac{\sqrt{2\pi}}{\sigma_y}\right)},
\end{equation}
with $y = R\theta$ the azimuthal coordinate along the ring, $A_{\rm gas}$ the gas overdensity contrast, $\sigma_y$ its azimuthal width, $\mathrm{Erf}$ the error function, and the normalisation is chosen so that the azimuthal average of the surface density would be its radial profile. In addition to the surface density parameters $A_{\rm gas}$ and $\sigma_y$, the analytical model of \citet{Birnstiel2013} depends also on the viscosity $\alpha$, the maximum grain size $a_{\rm max}$, the fragmentation velocity $v_{\rm frag}$ and the dust-to-gas volume ratio $\epsilon$. Then, by visually comparing the observed Stokes number measured from their discs with the model prediction, \citet{vdM2021} found that the best set of parameters describing their sample, including IRS\,48 and HD\,142527, was: $A_{\rm gas}=1.2$, $\sigma_y=10\degree$, $\alpha=10^{-3}$, $a_{\rm max}= 1~\rm mm$, $v_{\rm frag}=5 ~\rm m s^{-1}$ and $\epsilon=0.1$. However, they also specify the presence of degeneracies in the parameter space, for example between viscosity, dust-to-gas ratio and/or gas overdensity, or the use of either a grain size distribution with a fixed maximum grain size or the steady-state equilibrium model from \citet{Birnstiel2011}.

We follow the same approach to compute the emission azimuthal profile and reproduce our new HD\,34700\,A data, assuming a power law gas surface density radial profile $\Sigma_{\rm g}(R)=\Sigma_{\rm c}(R/R_{\rm c})^{-1}$ and evaluating Eq.~\ref{eq:azimuthal surf dens} at the radial location of the peak intensity maximum $R_{\rm peak}$. We assume the steady-state dust distribution from \citet{Birnstiel2011} for the dust population and we compute the optical depth $\tau=\sum_i \Sigma_i\kappa_{\mathrm{abs},i}$, with $\Sigma_i$ and $\kappa_{\mathrm{abs},i}$ the surface density and ``DSHARP"-mixture absorption opacity associated with grains of size $i$, taken from \citet{Birnstiel_2018}, and finally obtain the intensity profile of the emission $I_\nu = B_\nu(T_{\rm dust})(1-e^{-\tau})$, assuming a vertically isothermal disc with radial profile $T_{\rm dust}=T_{\rm c}(R/R_{\rm c})^{-0.5}$, with $T_{\rm c} = 20~\rm K$ at $R_{\rm c} = 100~\rm au$.

\begin{figure}
   \centering
   \includegraphics[width=\hsize]{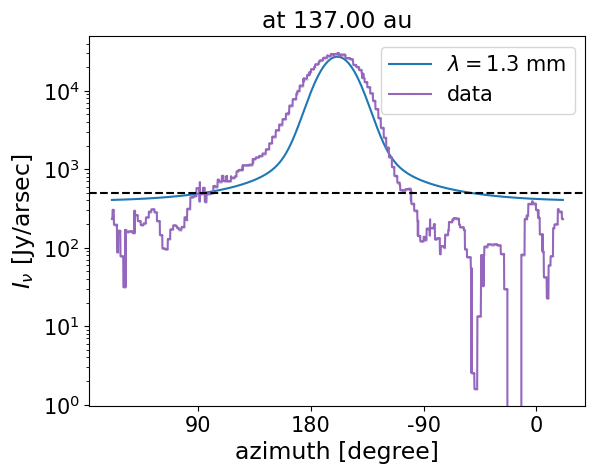}
      \caption{Comparison between azimuthal profile of the observed dust emission at the radial location of the peak intensity (purple line) and model intensity azimuthal profile (blue line) for the HD\,34700\,A system. The black dashed line corresponds to the azimuthal average of the observed intensity at $170~\rm au$.
              }
         \label{fig:Azimuthal model}
\end{figure}

In Figure~\ref{fig:Azimuthal model}, we show the observed azimuthal profile at the radial location of the peak intensity ($137~\rm au$) and the intensity profile of the model, obtained by manually changing the input parameters until visually reproducing the observed profile. Although the parameter space is highly degenerate, as discussed in \citet{vdM2021}, we can constrain some of the parameters under the assumption that the dust overdensity is caused by a vortex. In fact, vortices need low viscosity to be long lived \citep[$\alpha\lesssim10^{-4}-10^{-3}$;][]{Zhu2014,Bae2015, Rometsch_2021} and concentrate dust such that the dust-to-gas ratio is locally increased. As a result, we find that the observed azimuthal profile is best reproduced for the following choice of parameters: $A_{\rm gas}=1.1$, $\sigma_y=75\degree$, $\alpha=5\times10^{-4}$, $\Sigma_{\rm c}= 50~\rm g cm^{-2}$ at $R_{\rm c} = 1~\rm au$, $v_{\rm frag}=5 ~\rm m s^{-1}$ and $\epsilon=0.1$.

While this set of parameters results in a model matching the observed intensity azimuthal profile on a qualitative level, it should not be taken as a unique solution. Increasing (decreasing) the $\alpha$ viscosity produces a lower (higher) peak-to-valley ratio of the profile, while raising (reducing) either $\epsilon$ or $\Sigma_{\rm g}$ shifts the azimuthal profile upwards (downwards). The width of the azimuthal profile is proportional azimuthal width $\sigma_y$ of the gas bump, whereas increasing (decreasing) the overdensity contrast $A_{\rm gas}$ increments (decrements) the peak-to-valley ratio of the profile and produces a narrower (wider) profile. As a result, different combinations of the parameters might still reproduce the observed profile. However, with this analysis we do not aim to obtain a quantitative description of the observe profile, but to demonstrate that it is possible to reproduce it by invoking the dust trapping scenario. 

Although this model can reproduce qualitatively the observed intensity azimuthal profile of the peak emission for the HD\,34700\,A system, it does not recover the elongated trailing tail that we see in Fig.\ref{fig:Continuum gallery} and \ref{fig:Azimuthal model}. The asymmetric nature of the azimuthal profile might be indicative of dust segregation, with different sized dust grains trapped at different locations. This effect was shown in the simulations of \citet{Hammer_2019}, where they studied an elongated planet-induced vortex, featuring large grains concentrated close to the vortex centre and small grains circulating across the entirety of the vortex, producing an off centred peak emission and a skewed azimuthal profile. Multi-wavelength observations are needed to confirm the dust segregation scenario and further assess the vortex origin of the asymmetric emission.

\subsection{Origin of the asymmetry}
\label{sec:Vortex}

A possible explanation for the high contrast continuum asymmetries observed in Figure~\ref{fig:Continuum gallery} is gas anticyclonic vortices trapping dust in both the radial and azimuthal directions. Indeed, vortices have been suggested to be present in all the systems we consider in this work. 

For IRS\,48, \citet{vdM2013} reproduced the crescent by first simulating the gas distribution using FARGO3D, with a single sub-stellar companion carving the cavity for $1200~\rm orbits$, exciting RWI and producing a vortex. Then, starting from the resulting gas surface density, they computed the dust evolution and distribution with the model from \citet{Birnstiel2013}, reproducing the observed contrast. Then, from polarimetric ALMA observations, \citet{Yang2024} inferred that the dust grains in the overdensity of IRS\,48 ~are moderately settled, with a gas-to-dust scale height ratio of 2, and measured a turbulent $\alpha$ parameter in the range $[10^{-4}-5\times10^{-3}]$, consistent with the presence of a turbulent vortex. In HD\,142527,  \citet{Boehler2021} observed kinematic deviations from the gas Keplerian rotation in the $^{13}$CO and C$^{18}$O line emission velocity maps, consistent with a $\pm40~\rm au$, $200\degree$ large vortex, although a similar pattern might arise as a consequence of beam smearing in the presence of gas pressure gradients at the inner and outer edges of the disc. Multiple line, high resolution observations are needed to confirm either scenario. Finally, \citet{Stadler_2026} recently analysed high resolution ($0\farcs1$) ALMA observations of $^{12}$CO, $^{13}$CO and C$^{18}$O lines in the system HD\,34700\,A. They find tentative evidence of an elongated Doppler flip co-located with the continuum asymmetry in $^{13}$CO and C$^{18}$O, similarly to HD\,142527 ~\citep{Boehler2021}, and model it as a kinematic signature of a vortex \citep{Wolfer2025}.

These three systems host young ($<10~\rm Myr$) stars, which typically have a higher multiplicity fraction compared to main sequence stars \citep{Offner_2023}. Both HD\,34700\,A ~and HD\,142527 ~host a binary with a mass ratio $q>0.05$, which naturally carve an eccentric cavity \citep{Ragusa2017,Ragusa2020} with size of up to ($6 a_{\rm bin}$) ~\citep{Penzlin2024, Penzlin_2025, Sudarshan_2022}. However, with a binary separation of $0.69~\rm au$ for HD\,34700\,A ~\citep{Sterzik2005} and $10.8~\rm au$ for ~HD\,142527 \citep{Nowak2024}, the observed cavity sizes are too large to be explained by binary interaction alone, as shown by the simulations of HD\,142527 ~in \citet{Penzlin2024}.

Interestingly, a possible cause for the formation of a vortex is the presence of a planet carving the cavity \citep{Bae_2016, Lyra_2009, deValBorro2007, Hammer_2023}. Indeed, a massive companion would help the binaries of HD\,34700\,A ~and HD\,142527, and the star in IRS\,48 ~clear their large cavities, sustaining the RWI instability at its edge and triggering the formation of an anti-cyclonic vortex \citep{Hammer_2021}. The presence of a planet might also explain the asymmetrical nature of the crescent we see in Figure~\ref{fig:Continuum gallery} \citep{Hammer_2019}. Notably, \citet{Hammer_2017} have shown that slowly-growing planets trigger weaker but more elongated vortices, which, in combination with typical vortex lifetimes of $\lesssim1000$ local orbits \citep{Fung_2021,Rometsch_2021}, suggests that the object responsible for the observed vortex must not have migrated too far inwards from the current location of the vortex between its formation and today. Testing this scenario would require follow-up observations aiming to detect a companion within the disk cavity.

Infalling streamers hitting the disc are another mechanism that can produce vortices \citep{Bae2015, Kuznetsova_2022}. While IRS\,48 ~and HD\,142527 ~show no evidence of interactions with the environment, \citet{Stadler_2026} have suggested the presence of infalling streamers on the HD\,34700\,A ~disc from spectral line analysis of $^{12}$CO. The strongest deviations from the disc rotation, probably associated to the location where the streamer hits the disc, are located in the vicinity of the dust crescent. It is unclear if the asymmetry in HD\,34700\,A is due to the infall, the presence of an undetected massive companion carving the cavity or a combination of the two. 

HD\,142527 is unique in that the asymmetry is best fit by two Gaussian curves in azimuth with our model. This configuration could be explained by either two recently-formed Rossby-wave-instability vortices that are in the process of coalescing, or the superposition of a vortex and the bright emission near the apocentre of an eccentric cavity \citep[e.g.,][]{Penzlin2024,Calcino_2019}. The former scenario is highly unlikely due to its narrow time window. On the other hand, a massive companion could excite both an eccentric cavity \citep{Kley_2006,Tanaka_2022} and a massive vortex on its edge, making it a favourable formation scenario in the absence of environmental factors such as infall.

\subsection{Dust distribution around a massive vortex}
\label{sec:hydro-vortex}

\begin{figure}
    \centering
    \includegraphics[width=\columnwidth]{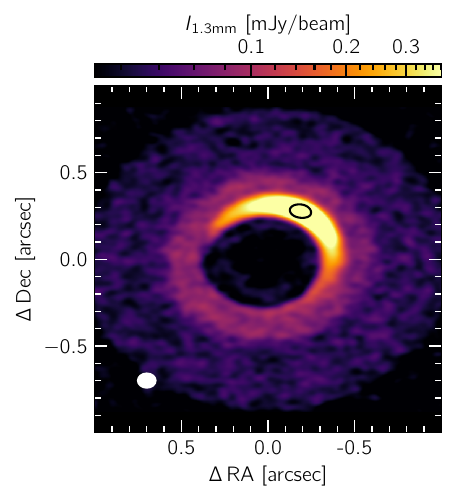}
    \caption{Synthetic observation of an HD\,34700\,A analogue using a multi-fluid hydrodynamical simulation of a vortex. The emission is optically thin with the exception of the centre of the vortex (black circle after convolving). A slight tail is visible on the simulated vortex, similar to the observations of HD\,34700\,A and IRS\,48.}
    \label{fig:vortex-model}
\end{figure}

In the previous paragraph we discussed several possible scenarios for the origin of the observed dust asymmetries, assuming they are vortices. Here, we would like to investigate the distribution of dust grains across the vortex, including the degree of asymmetry with respect to its centre, regardless of the origin of the vortex.

We model an HD\,34700\,A analogue using the hydrodynamics code \texttt{PLUTO} \citep{Mignone_2007} with a vertically-integrated, cylindrical ($\{R,\phi\}$) grid extending between 0.4--2.5\,$R_0$ in radius with $R_0=150$\,au, and spanning the full $2\pi$ in azimuth, with $N_R\times N_\phi = 512\times1024$ cells and logarithmic spacing in radius. We utilize second-order spatial and temporal accuracy, the HLLC solver \citep{Toro_1994}, a locally isothermal equation of state, the dust fluid module described in \citet{Ziampras_2025}, and the FARGO method to both improve numerical accuracy and drastically accelerate computations \citep{Masset_2000,Mignone_2012}. The central object is modelled as a single star with $M_\star=4\,\text{M}_\odot$.

The numerical model is initialized with a disc of gas and dust that follow radial power-law profiles with $\Sigma_\text{g}=1\,\text{g cm}^{-2}\,(R/R_0)^{-1}$ and $\Sigma_\text{d}=0.01\,\Sigma_\text{g}$, including a cavity interior to $R_0$ similar to \citet{Sudarshan_2022}, and an aspect ratio profile $h=0.15\,(R/R_0)^{1/4}$. Finally, we include viscous stress following \citet{Shakura_Sunyaev_1973} with $\alpha=10^{-4}$ and a dust diffusion prescription following \citet{Weber_2019}.

We include six dust fluids with fixed grain sizes $a_\text{d}=\{10^{-4}, 0.001, 0.003, 0.01, 0.03, 0.1\}$\,cm and bulk densities of $\tilde\rho=2.08\,\text{g cm}^{-3}$, coupled to the gas in the Epstein drag regime. The initial dust fluid densities are weighed such that $\Sigma_{\text{d},i} \propto a_{\text{d},i}^{q+4}$ with $q=-3.5$, corresponding to an MRN distribution \citep{Mathis_1977}.

Our profile for $h$ is quite in line with expectations beyond 100\, au for typical disc parameters, even if possibly rather unrealistic for a passively irradiated disc with the particular stellar parameters for HD\,34700\,A. Nevertheless, given that we are aiming for a generic, proof-of-concept model, and that the cavity edge in HD\,34700\,A is anyway likely directly exposed to stellar irradiation (thus increasing the temperature and therefore $h$ to values around 0.1--0.12), this choice is justified for our purposes.

With these initial conditions, the cavity edge at $\sim\!R_0$ is unstable to the Rossby-wave instability \citep{Lovelace1999} and should collapse to a vortex. To form a massive vortex, we added a radially Gaussian distribution of gas and dust with a FWHM of $2 H_0 = 0.3\,R_0$ centred at $R_0$, and seeded their velocity fields with noise of up to 1$\%$ of the local sound speed before integrating for 100 orbits at $R_0$.

Immediately as the simulation begins, the cavity edge collapses into several vortices that coalesce into a single massive vortex over the first $\sim20$\,orbits. By the end of the simulation, a large, quasi-stable vortex has been established, with dust species of different sizes trapped at varying degree around it.

To produce the synthetic image at $\nu=230~\rm GHz$ ($\lambda=~1.3$\,mm) we compute the intensity at that frequency using
\begin{equation}
    \label{eq:intensity}
    I_\nu(\bm{R}) = B(\nu,T)\,\left(1-e^{-\tau_\nu}\right),\quad\tau_\nu(\bm{R}) = \sum\limits_{i}\kappa_{\nu,i}\Sigma_{\text{d},i}(\bm{R}),
\end{equation}
which we then convolve with a beam of angular size $0\farcs114\times0\farcs094$ and layer with Gaussian noise convolved with the same beam and with an RMS of 7\,$\mu$Jy/beam. The result is shown in Fig.~\ref{fig:vortex-model}, using the same spatial extent and colorbar stretch as Fig.~\ref{fig:Continuum gallery}. We find that the synthetic image largely reproduces the features found around the cavity in the HD\,34700\,A system, with the emission around the vortex showing a (mild) asymmetrical tail (bottom right quadrant in the figure).

We note that the emission in our model can be rescaled to arbitrary radial scales and distances, provided that the dynamics of the gas and dust (related to $h$ and the Stokes number $\text{St}\propto\tilde{\rho}a_\text{d}/\Sigma_\text{g}$) are not changed. As a result, this model does not necessarily constrain the exact gas or dust properties around HD\,34700\,A beyond highlighting the requirement for large grains trapped in the vortex, which was already inferred in Sect.~\ref{sec:Dust trapping}. Nevertheless, our model shows that non-axisymmetric emission with a large azimuthal extent can be easily (re)produced with a massive vortex, a structure that can naturally form in protoplanetary discs.

\subsection{Disc eccentricity}
\label{sec:Disc eccentricity}

Our \texttt{galario} models assume circular rings and  arcs with no eccentricity. \citet{Yang2023} measured a high eccentricity of $0.27$ in the dust ring of IRS\,48, clearly motivated by the evident offset of the star with respect of the ring centre. In Figure~\ref{fig:Continuum gallery} we mark the centre of the disc corrected with the \texttt{galario} offset, which is displaced with respect to the inner disc emission that surrounds the central stars. This can be indicative of a non-zero eccentricity of the dust ring also in HD\,34700\,A ~and HD\,142527, which is more evident in the polar maps in Figure~\ref{fig:Azimuthal plots}. Indeed, the de-projected rings would appear as a straight horizontal line in these plots if they were circular, but instead the ring and the asymmetry peak at different radii, with the tail of the asymmetry connecting them spanning different radii. Additionally, azimuthal profiles taken at the same radial distance inward and outward from the peak intensity (see right column of Figure~\ref{fig:Azimuthal plots}, comparing orange with red profiles and blue with purple profiles) are different from each other. This can be due to the possible eccentricity of the rings, or indicative of an asymmetric morphology of the rings and overdensities along the radial direction, or a combination of the two effects. Using numerical 3D hydrodynamical simulations, \citet{Price_ea_2018} showed that an eccentric circumbinary ring can be the result of an eccentric binary orbit. Later on, numerical studies from \citet{Ragusa2020, Penzlin2024} demonstrated that circumbinary disc can develop non zero eccentricity also in the case of a binary with a circular orbit, or as a consequence of the thermodynamical properties of the disc \citep{Penzlin_2025}.

In order to measure the eccentricity of the discs, we expand Eqs.\eqref{eq:Intensity HD34700}-\eqref{eq:Intensity HD142527} to include eccentric rings and arcs. We parameterize the radial centre using:

\begin{equation}\label{eq:ellipse}
    r = a\frac{1-e^2}{1-e\cos(\phi - \omega)},
\end{equation}
where $a$ is the semi-major axis of the ellipse, $e$ its eccentricity and $\omega$ the argument of periastron. This parametrization assumes that the coordinate system is centred on the focus of the ellipse. We fix the eccentricity to be the same for all the ring and arc components in each disc model, and start the fitting procedure using these new parameters. We use the best-fit parameters we obtained for the circular models as initial guesses for the \texttt{galario} runs with the eccentric models, using the same setup we introduced in Sec.~\ref{sec:Methods}.

The HD\,34700\,A model converges to a solution with $a = 529^{+3}_{-3}~\rm mas$, $e = 0.109^{+0.005}_{-0.004}$, $\omega = 105^{+7}_{-7}~ \rm deg$, $\Delta\rm RA = 99.6^{+0.6}_{-0.3}~\mu as$, $\Delta\rm Dec = 19^{+6}_{-5}~\mu as$ and the remaining parameters compatible within $3\sigma$ with those of the circular model. In particular, we find an inclination $i=35.3^{+0.6}_{-0.5}~\rm deg$ and position angle $PA=90.0^{+0.5}_{-0.6}$, in remarkable agreement with the geometrical parameters of the circular model. We note that in this case we use wide priors for the geometrical parameters since the beginning of the fitting procedure. In the cases of IRS\,48 and HD\,142527, the eccentric \texttt{galario} models do not reach convergence. This is possibly caused by the presence of the asymmetry, which is radially wider compared to the ring and obscures the cavity edge morphology, making it compatible with a wide range of parameters. This degeneracy prevents the MCMC procedure from converging on a well-constrained parameter set. The IRS\,48 and HD\,142527 datasets are more affected than HD\,34700\,A as their asymmetries are azimuthally wider.

In Appendix~\ref{App: Eccentric model} we show the model and residuals images of the eccentric model, and compare them with the circular case. The offset of the central component in the eccentric model is possibly caused by the presence of emission between the inner disc component and the asymmetry. It is interesting to note that the residuals in both the eccentric and circular models do not show any substantial difference. This suggests that there is a degeneracy between the circular and eccentric model, and it is currently unclear if the disc around HD\,34700\,A is circular or eccentric.

\section{Conclusions}
\label{sec:Conclusions} 

In this paper we present new high resolution ($0\farcs11$) ALMA band 6 ($1.3~\rm mm$) continuum observations of the system HD\,34700\,A. We compare its emission to two other systems featuring a similar morphology, IRS\,48 and HD\,142527, and perform visibility modelling of the new ALMA data, while also improving existing models for the other two systems, using the code \texttt{galario}. Our results are summarised as follows:

\begin{itemize}
    \item In the continuum data of HD\,34700\,A we observe a cavity and resolve an asymmetric ring with a de-projected radius of $0\farcs53$ ($186~\rm au)$ and a peak intensity contrast of 62, making HD\,34700\,A the transition disc with the second most prominent continuum asymmetry detected by ALMA to date. The dust overdensity is asymmetric, with a tail trailing with respect to the rotation of the gas, and peaks at $0\farcs39$ ($137~\rm au$), with a PA of $227\degree$ and an azimuthal extent of more than $180\degree$.
    \item Our \texttt{galario} model for HD\,34700\,A is in remarkable agreement with the data, with residuals of up to $7.4\sigma$ localised only in the region of the asymmetric tail. Based on our model, we measure an azimuthal FWHM of $43.2\degree$ for the asymmetry. The central emission is offset compared to the disc centre corrected with the \texttt{galario} offset, suggesting that the ring is eccentric. We improve the existing \texttt{galario} model for HD\,142527 \citep{vdM2021}, retrieving the double peaked structure of the ring. In this case we observe residuals of up to $20.1\sigma$ ($6.2\%$ of peak intensity) all across the ring, an improvement by a factor $\sim6$ compared to the previous model, implying that the morphology is more complex than the Gaussian prescription in our model. For IRS\,48, instead, we obtain a model similar to that of \citet{vdM2021}, failing to recover the ring emission. We measure residuals of up to $96\sigma$ and $-150\sigma$, corresponding to $8.6\%$ and $13.5\%$ of the peak intensity, respectively.  
    \item We compare the continuum emission of the three transition discs with the most asymmetric rings. All systems have a large dust cavity that cannot be explained by binary-disc interaction alone, an inner emission offset with respect to the disc centre, indicative of an eccentric ring, and a high contrast overdensity with an asymmetric tail trailing with respect to the disc's rotation. The high azimuthal contrast of the asymmetries rules out the orbit clustering of eccentric cavities scenario \citep{Ataiee2013}. A vortex has been proposed to explain the origin of the dust asymmetry, through kinematic analysis for HD\,34700\,A \citep{Stadler_2026} and HD\,142527 \citep{Boehler2021}, and dust polarization measurements for IRS\,48 \citep{Yang2024}.
    \item We measure the eccentricity of the ring around HD\,34700\,A, adapting our visibility models to fit eccentric rings and arcs, finding an eccentricity of $0.109$. However, the model show a degeneracy with the circular solution, thus our estimates do not confirm that the system is eccentric.
    \item Under the vortex assumption, we reproduce the observed azimuthal profile of the emission using the analytical dust model of \citet{Birnstiel2013}, by manually changing the input parameters. However, we stress that this solution might not be unique, as the parameter space is highly degenerate, and serves the purpose of providing a qualitative idea for the physical parameters associated with the possible dust trapping vortex.
    \item A hydrodynamic model of a massive vortex yields qualitatively similar features to those found in HD\,34700\,A and, to an extent, IRS\,48, in that the emission around the vortex shows a mild asymmetry between the leading and trailing sides. In combination with the above point, this suggests that the vortex scenario is indeed a plausible explanation for these features.
\end{itemize}

We have shown that the observed asymmetry can be consistent with the dust trapping scenario. However, it is currently unclear which physical mechanism is causing the observed azimuthal asymmetric rings in the discs of HD\,34700\,A, IRS\,48 and HD\,142527. While the presence of a vortex might explain the observed signatures, and it has been tentatively proposed for all the systems we consider, an alternative explanation might lie in the presence of a gas overdensity induce by eccentric cavities due to binary companions \citep{Shi2012, Ragusa2017, Miranda_2017}. In order to differentiate between the two scenarios, higher spatial resolution ($\lesssim0\farcs1$) observations are needed to detect the rotational kinematic pattern of a vortex. At the same time, multi-wavelength observations of the dust continuum emission would constrain the distribution of dust grains with different sizes, which are expected to show a shift in the spatial location of the peak emission in the presence of a vortex \citep{Cazzoletti_2018}.

\begin{acknowledgements}
         DF acknowledges Clement Baruteau, Aleksandra Kuznetsova, Anibal Sierra, and Daniel Price for the helpful scientific discussions. DF also acknowledges Milou Temmink and Haifeng Yang for sharing the calibrated data of HD 142527 and IRS 48. MB acknowledges Marco Tazzari for the development of \texttt{Galario}. This project has received funding from the European Research Council (ERC) under the European Union’s Horizon 2020 research and innovation programme (PROTOPLANETS, grant agreement No. 101002188). S.F. is funded by the European Union (ERC, UNVEIL, 101076613) and acknowledges financial contribution from PRIN-MUR 2022YP5ACE. ER acknowledges support from the European Union (ERC Starting Grant DiscEvol, project number 101039651). Views and opinions expressed are, however, those of the author(s) only and do not necessarily reflect those of the European Union or the European Research Council. Neither the European Union nor the granting authority can be held responsible for them.  
        
        This paper makes use of the following ALMA data: ADS/JAO.ALMA\#2013.1.00305.S, ADS/JAO.ALMA\#2019.1.01059.S, ADS/JAO.ALMA\#2022.1.00760.S . ALMA is a partnership of ESO (representing its member states), NSF (USA) and NINS (Japan), together with NRC (Canada), NSTC and ASIAA (Taiwan), and KASI (Republic of Korea), in cooperation with the Republic of Chile. The Joint ALMA Observatory is operated by ESO, AUI/NRAO and NAOJ.

\end{acknowledgements}

\bibliographystyle{aa}
\bibliography{aa58212-25}

\begin{appendix} 

\onecolumn

\section{Observations}\label{App: Observations}

\begin{table*}[h]
\caption{Summary of the data used in this work. }            
\label{table:observation_summary}      
\centering                          
\footnotesize
\renewcommand{\arraystretch}{1.5}
\begin{tabular}{l c c c c c c r}        
\hline\hline                
Label & Project ID & Observation Date & Baselines & Frequency & Resolution & Max. Scale & References\\
 &    &  & [m] & [GHz]& [arcsec] & [arcsec] \\
\hline                        
HD\,34700\,A & 2022.1.00760.S  & 2022 Oct 09  &  33–214  &  344–355   & 0.72  & 8.57 &  \citet{Stadler_2026}\\
        & 2022.1.00760.S  & 2023 May 05 - Jun 02  &  158–1417  &  346–355   &  0.12 & 2.30 &  \citet{Stadler_2026}\\

\hline     
\end{tabular}
\end{table*}

In Table~\ref{table:observation_summary} we report the observing log associated to the ALMA datasets presented for the first time in this paper.

\section{Visibility Modelling}\label{App: Visibility Modelling}

    \begin{table*}[h]
    \caption{\texttt{galario} best fit parameters}             
    \label{table:galario results}      
    \centering      
    \footnotesize
    \renewcommand{\arraystretch}{1.5}
    \begin{tabular}{c c c c c c c c c c c}     
    \hline\hline \\     
                          
    Dataset & $\log_{10}f_0$ & $x_0$ & $y_0$ & $\sigma_{r,0}$ & $\log_{10}f_1$ & $r_1$ & $\sigma_{r,1}$ & $\log_{10}f^a_1$ & $r^a_1$ \\ 
     & [Jy sr$^{-1}$] & [mas] & [mas] & [mas] & [Jy sr$^{-1}$] & [mas] & [mas] & [Jy sr$^{-1}$] & [mas] \\ \\
    \hline\\                    
       HD\,34700\,A & $8.10_{-0.03}^{+0.03}$ & $-$ & $-$ & $130_{-6}^{+6}$ & $8.15_{-0.01}^{+0.01}$ & $525_{-2}^{+2}$ & $84_{-3}^{+3}$ & $10.148_{-0.002}^{+0.002}$ & $394_{-2}^{+2}$ \\  
       IRS\,48 & $-$ & $-$ & $-$ & $-$ & $8.052_{-0.007}^{+0.007}$ & $719_{-4}^{+4}$ & $172_{-3}^{+3}$ & $11.04432_{-0.00008}^{+0.00008}$ & $420.5_{-0.3}^{+0.3}$ \\
       HD\,142527 & $10.2_{-0.1}^{+0.1}$ & $94_{-3}^{+3}$ & $-159_{-3}^{+2}$ & $34_{-5}^{+4}$ & $10.950_{-0.005}^{+0.004}$ & $1122.29_{-0.03}^{+0.03}$ & $3.22_{-0.03}^{+0.03}$ & $10.6383_{-0.0008}^{+0.0008}$ & $1202.4_{-0.3}^{+0.3}$ \\ \\
    \hline\hline \\          
    Dataset & $\sigma^a_{r,1}$ & $\theta^a_1$ & $\sigma^a_{\theta,1\rm l}$ & $\sigma^a_{\theta,1\rm r}$ & $\log_{10}f^a_2$ & $r^a_2$ & $\sigma^a_{r,2}$ & $\theta^a_2$ & $\sigma^a_{\theta,2\rm l}$ \\ 
     & [mas] & [deg] & [deg] & [deg] & [Jy sr$^{-1}$] & [mas] & [mas] & [deg] & [deg] &  \\
    \hline \\                   
       HD\,34700\,A & $27.8_{-0.2}^{+0.2}$ & $327.7_{-0.3}^{+0.3}$ & $23.5_{-0.1}^{+0.1}$ & $13.2_{-0.1}^{+0.1}$ & $-$ & $-$ & $-$ & $-$ & $-$ & \\  
       IRS\,48 & $72.13_{-0.02}^{+0.02}$ & $197.34_{-0.04}^{+0.05}$ & $28.37_{-0.03}^{+0.03}$ & $29.49_{-0.02}^{+0.02}$ & $-$ & $-$ & $-$ & $-$ & $-$ & \\
       HD\,142527 & $141.1_{-0.2}^{+0.2}$ & $298.06_{-0.04}^{+0.04}$ & $18.56_{-0.04}^{+0.04}$ & $51.14_{-0.04}^{+0.04}$ & $10.7017_{-0.0001}^{+0.0001}$ & $1318.1_{-0.2}^{+0.2}$ & $184.08_{-0.06}^{+0.06}$ & $227.23_{-0.02}^{+0.02}$ & $29.74_{-0.02}^{+0.02}$ & \\ \\
    \hline\hline \\ 
    Dataset & $\sigma^a_{\theta,2\rm r}$ & $i$ & $\rm PA$ & $\Delta \rm RA$ & $\Delta \rm Dec$ & \\ 
     & [deg] & [deg] & [deg] & [mas] & [mas] \\ \\
    \hline \\                   
       HD\,34700\,A  & $-$ & $35.1_{-0.1}^{+0.1}$ & $92.9_{-0.3}^{+0.3}$ & $-0.048_{-0.001}^{+0.001}$ & $0.014_{-0.001}^{+0.001}$ &  \\  IRS\,48 & $-$ & $50.996_{-0.005}^{+0.003}$ & $100.12_{-0.04}^{+0.04}$ & $83.8_{-0.3}^{+0.3}$ & $107.8_{-0.2}^{+0.2}$ & \\ 
       HD\,142527 & $61.0_{-0.1}^{+0.1}$ & $29.079_{-0.004}^{+0.004}$ & $182.96_{-0.03}^{+0.03}$ & $-38.1_{-0.2}^{+0.2}$ & $-154.6_{-0.2}^{+0.2}$ & \\ \\
    \hline
    \end{tabular}
   
    \end{table*}

    \begin{figure*}[h!]
        \centering 
        \begin{subfigure}[t]{0.32\columnwidth}
        \centering
        \includegraphics[width=\columnwidth]{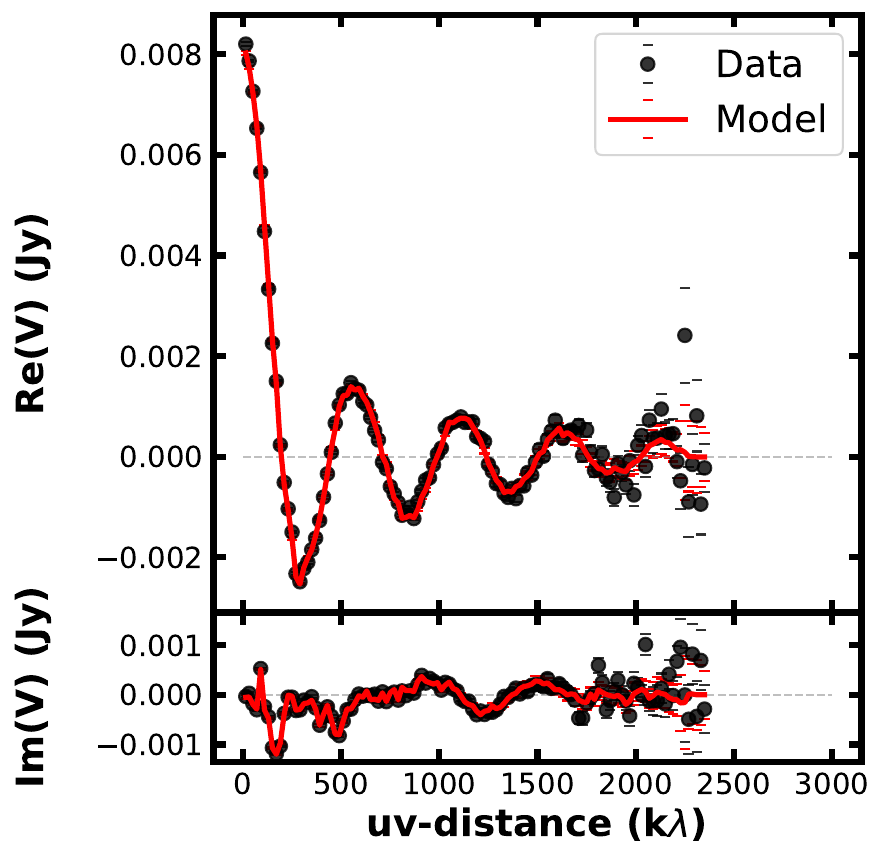}
        \caption{
              }
         \label{fig:HD34700 Visibility Profile}
         \end{subfigure}
         \hfill
         \centering 
        \begin{subfigure}[t]{0.32\columnwidth}
        \centering
        \includegraphics[width=\columnwidth]{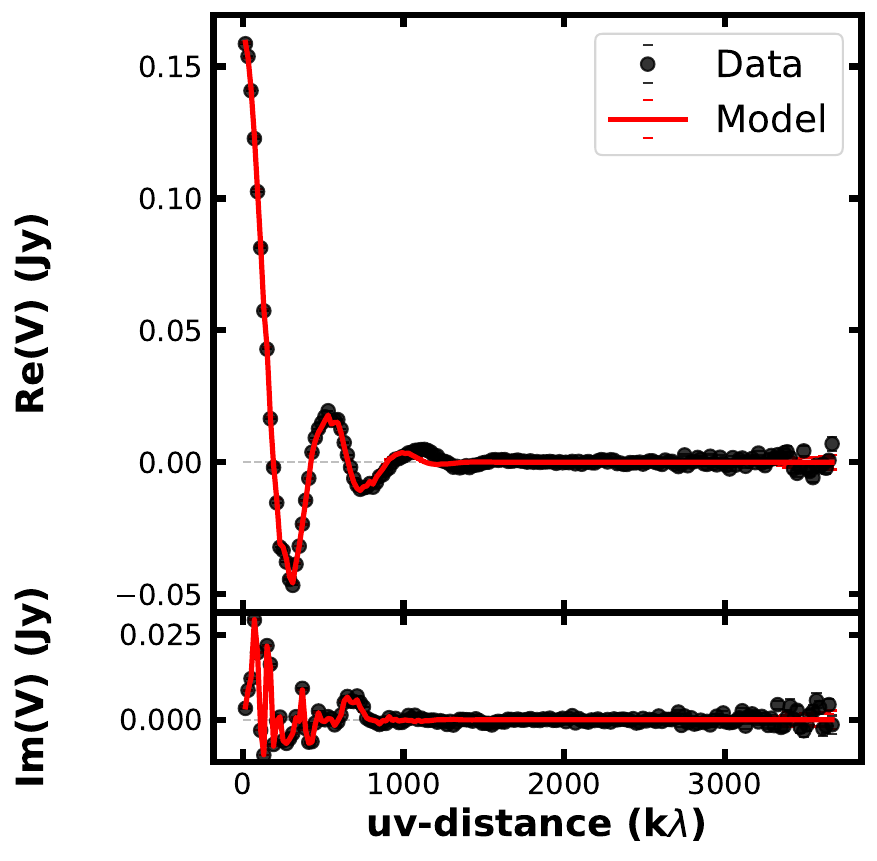}
        \caption{
              }
         \label{fig:IRS48 Visibility Profile}
         \end{subfigure}
         \hfill
        \begin{subfigure}[t]{0.32\columnwidth}
        \centering
        \includegraphics[width=\columnwidth]{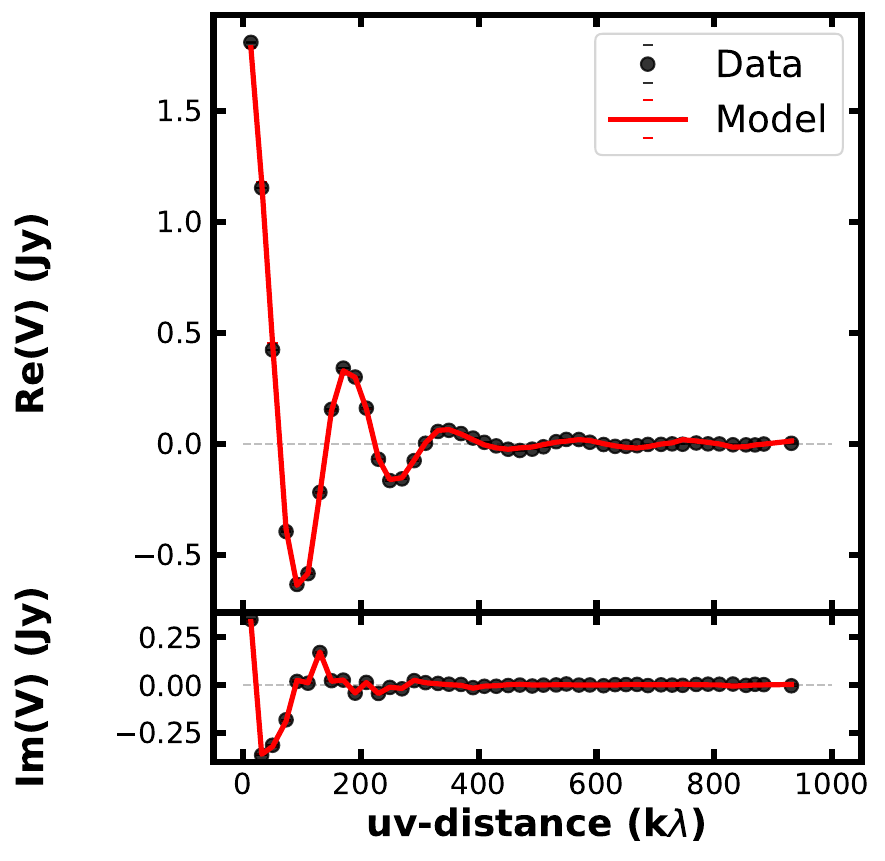}
        \caption{
              }
         \label{fig:HD142527 Visibility Profile}
         \end{subfigure}
    \caption{Real (top panel) and imaginary (bottom panel) parts of the re-centred and de-projected visibilities azimuthally averaged in bins with width of 20 $\rm k\lambda$ as a function of the de-projected baseline length for HD\,34700\,A (left), IRS\,48 (middle) and HD\,142527 (right). The black points represent the visibility data points, while the red solid line corresponds to the \texttt{galario} best fit model.
              }
    \label{fig:Visibility profiles}
   \end{figure*}

    In Table~\ref{table:galario results} we summarise the best fit parameters resulting from our visibility modelling of HD\,34700\,A, IRS\,48, HD\,142527 observations. We have adopted the following priors: 
    
    \noindent$HD\,34700\,A = \{\log_{10}f_0=[1.0, 10.0]$, $r_0=[0.0, 0.15]\rm~arcsec$,$\log_{10}f_1=[1.0, 10.0]$, $r_1=[0.0, 1.0]\rm~arcsec$, $\sigma_{r,1}=[0.0, 0.2]\rm~arcsec$, $\log_{10}f^{\rm a}_1=[1.0, 11.0]$,  $r^{\rm a}_1=[0.0, 1.0]\rm~arcsec$, $\sigma^{\rm a}_{r,1}=[0.0, 0.2]\rm~arcsec$, $\theta^{\rm a}_1=[0.0, 360.0]\rm~arcsec$, $\sigma^{\rm a}_{\theta,l1}=[0.0, 180.0]\rm~arcsec$, $\sigma^{\rm a}_{\theta,r1}=[0.0, 180.0]\rm~arcsec$, $i=[35.0, 37.0]\rm~deg$, $\rm PA=[92.5, 93.5]\rm~deg$, $\rm\Delta Ra=[-0.1, 0.1]\rm~arcsec$, $\rm\Delta Dec=[-0.1, 0.1]\rm~arcsec\}$, 
    
    \noindent$IRS\,48 = \{\log_{10}f_1=[1.0, 11.0]$, $r_1=[0.4, 1.0]\rm~arcsec$, $\sigma_{r,1}=[0.0, 0.2]\rm~arcsec$, $\log_{10}f^{\rm a}_1=[1.0, 12.0]$,  $r^{\rm a}_1=[0.4, 0.6]\rm~arcsec$, $\sigma^{\rm a}_{r,1}=[0.0, 0.2]\rm~arcsec$, $\theta^{\rm a}_1=[0.0, 360.0]\rm~arcsec$, $\sigma^{\rm a}_{\theta,l1}=[0.0, 180.0]\rm~arcsec$, $\sigma^{\rm a}_{\theta,r1}=[0.0, 180.0]\rm~arcsec$, $i=[49.0, 51.0]\rm~deg$, $\rm PA=[99.0, 101.0]\rm~deg$, $\rm\Delta Ra=[0.0, 0.5]\rm~arcsec$, $\rm\Delta Dec=[0.0, 0.5]\rm~arcsec\}$,
    
    \noindent$HD\,142527 = \{\log_{10}f_0=[1.0, 12.0]$, $x_0=[-0.2,0.2]$, $y_0=[-0.2,0.2]$, $r_0=[0.0, 0.2]\rm~arcsec$,$\log_{10}f_1=[1.0, 12.0]$, $r_1=[0.0, 2.0]\rm~arcsec$, $\sigma_{r,1}=[0.0, 0.8]\rm~arcsec$, $\log_{10}f^{\rm a}_1=[1.0, 12.0]$,  $r^{\rm a}_1=[0.0, 2.0]\rm~arcsec$, $\sigma^{\rm a}_{r,1}=[0.0, 1.0]\rm~arcsec$, $\theta^{\rm a}_1=[0.0, 360.0]\rm~arcsec$, $\sigma^{\rm a}_{\theta,l1}=[0.0, 180.0]\rm~arcsec$, $\sigma^{\rm a}_{\theta,r1}=[0.0, 180.0]\rm~arcsec$, $\log_{10}f^{\rm a}_2=[1.0, 12.0]$,  $r^{\rm a}_2=[0.0, 2.0]\rm~arcsec$, $\sigma^{\rm a}_{r,2}=[0.0, 1.0]\rm~arcsec$, $\theta^{\rm a}_2=[0.0, 360.0]\rm~arcsec$, $\sigma^{\rm a}_{\theta,l2}=[0.0, 180.0]\rm~arcsec$, $\sigma^{\rm a}_{\theta,r2}=[0.0, 180.0]\rm~arcsec$, $i=[20.0, 40.0]\rm~deg$, $\rm PA=[150.0, 250.0]\rm~deg$, $\rm\Delta Ra=[-0.3, 0.3]\rm~arcsec$, $\rm\Delta Dec=[-0.3, 0.3]\rm~arcsec\}$.

We choose narrow priors for the PA and inclination of HD\,34700\,A in order to be consistent with the analysis of \citet{Stadler_2026}, using the values resulting from their $^{13}$CO \texttt{discminer} fit as initial guesses to help achieve convergence. We follow the same approach for IRS\,48, using the inclination and PA reported in \citet{Yang2023}. At a later stage, we check that allowing for wider priors does not significantly change the geometrical parameters.
    
    In Figure~\ref{fig:Visibility profiles} we show the azimuthally averaged visibility profiles for HD\,34700\,A, IRS\,48 and HD\,142527 (black dots) and the best fit \texttt{galario} model (red solid line).

\section{Eccentric model}\label{App: Eccentric model}

 \begin{figure}[h!]
        \centering 
        \begin{subfigure}[t]{0.49\columnwidth}
        \centering
        \includegraphics[width=\columnwidth]{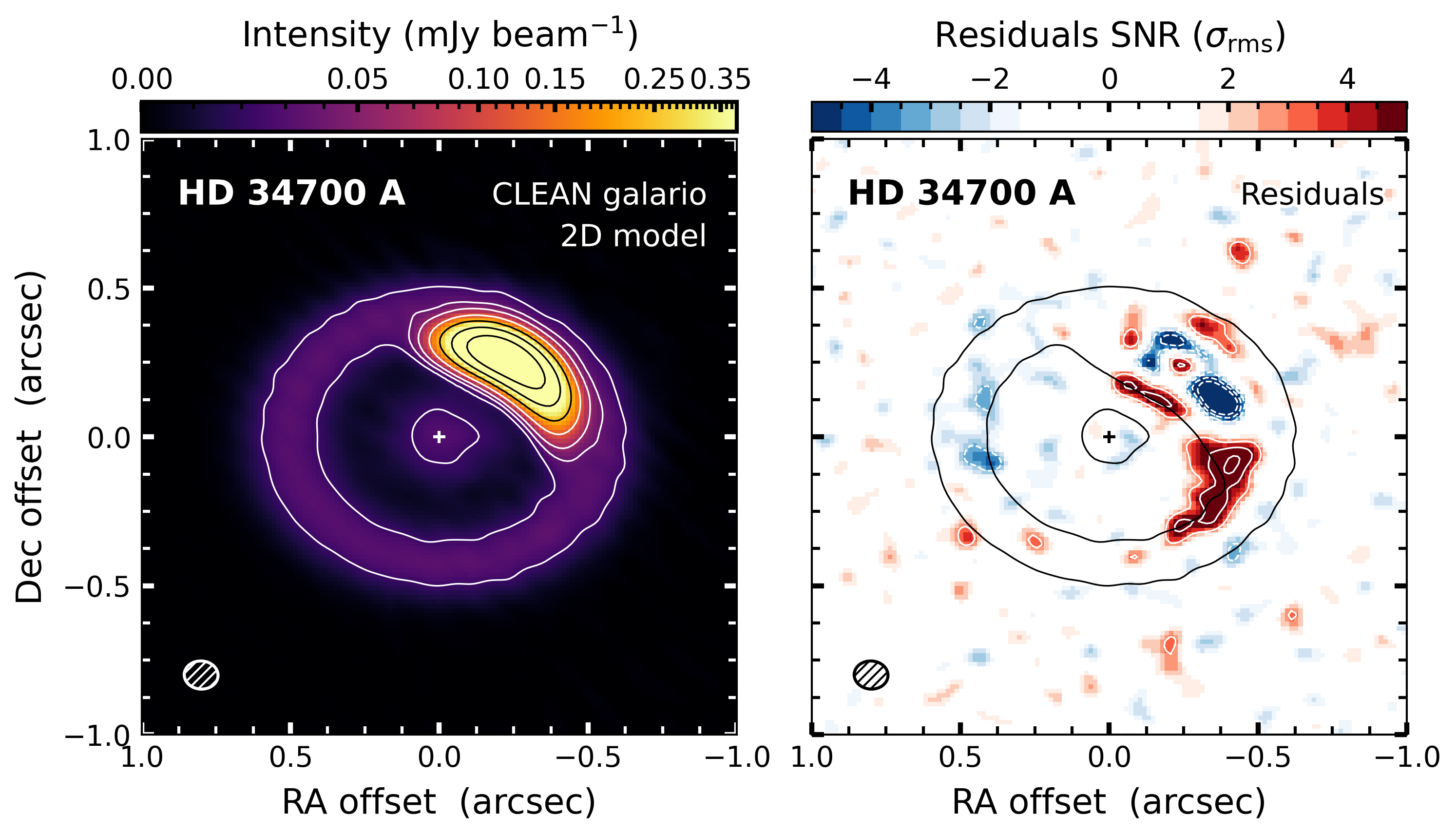}
        \caption{
              }
         \label{fig:Model and res circular}
         \end{subfigure}
         \hfill
         \centering 
        \begin{subfigure}[t]{0.49\columnwidth}
        \centering
        \includegraphics[width=\columnwidth]{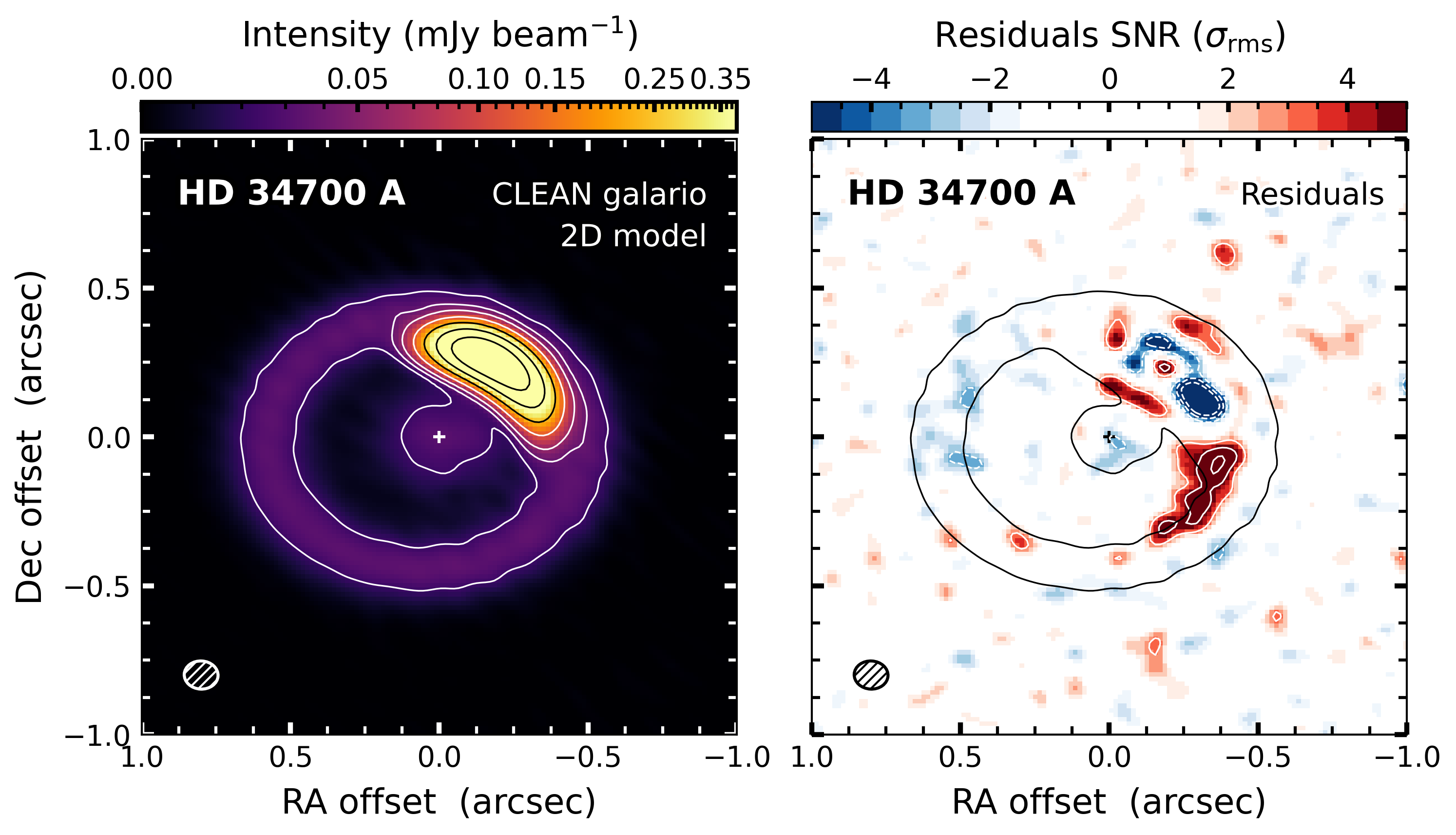}
        \caption{
              }
         \label{fig:Model and res elliptic}
         \end{subfigure}
    \caption{CLEANed images of the \texttt{galario} model (left panel) and residuals (right panel) for the circular (a) and eccentric (b) models of HD\,34700\,A, same as middle and right columns of Fig.~\ref{fig:Summary plots}.
              }
    \label{fig:Elliptic comparison}
   \end{figure}

In Fig.~\ref{fig:Elliptic comparison} we compare the \texttt{galario} models and residuals of the HD\,34700\,A system, assuming a circular model (same as middle and right columns of Fig.~\ref{fig:Summary plots}), with those obtained assuming an eccentric model, as explained in Sec.~\ref{sec:Disc eccentricity}.

\end{appendix}

\end{document}